\documentclass[12pt,pra,aps,amssymb,amsfonts,amsmath,tightenlines]{revtex4}
\usepackage[caption = false]{subfig}
\usepackage{dsfont}
\usepackage{graphicx}
\usepackage{amssymb,amsfonts,amsthm}
\usepackage{color}
\usepackage{bm,float}

\newcommand{\half}{\mbox{$\textstyle \frac{1}{2}$}}

\newcommand{\re}{\mbox{$\rm e$}}

\begin{document}

\title{The adaptive nature of confirmation bias}

\author{
Dorje~C.~Brody$^{1,2}$, Karl~J.~Friston$^{3}$, 
Bernhard~K.~Meister$^{4}$, Emmanuel~M.~Pothos$^{5}$}

\affiliation{
$^{1}$School of Mathematics and Physics, University of Surrey, 
Guildford GU2 7XH, UK \\  
$^{2}$Department of Mathematics, Imperial College London, London SW7 2BZ, UK \\ 
$^{3}$Queen Sq. Institute of Neurology, Dept. Imaging Neuroscience. University College London, London WC1N 3AR, UK  \\ 
$^{4}$K\"uhltal 28, 2120 Wolkersdorf, Austria \\ 
$^{5}$Department of Psychology, City St. George's, University of London, London EC1R 0JD, UK 
}

\date{\today}
\begin{abstract}
\noindent 
In this paper, the phenomenon generally classified as confirmation bias is formulated on the space of square-root probabilities (or equivalently, using the structures of quantum probability). In this framework, observations are modelled by matrices, rather than random variables on a probability space. In the problem of binary hypothesis testing, an optimal evidence choice minimises the expected error probability. We show that the resulting optimal choice of evidence leads to a confirmation bias, thus revealing a surprising aspect of rationality that encompasses confirmation bias.  Specifically, in sequential evidence sampling, the implicit optimality leads to two remarkable evolutionary advantages, namely, (a) the decision maker requires only the smallest memory capacity, and (b) the error probability can be reduced exponentially in sample size. A complementary approach based on the framework of active inference --- where the decision maker seeks evidence that provides maximum information --- is then considered. The resulting optimal evidence is shown to agree with the one obtained by minimising error probability. Our framework provides an easy-to-implement protocol for an active quantum inference, whereby the optimal evidence choice for making an inference is sought over the space of matrices.
\vspace{-0.2cm}
\\
\end{abstract}

\maketitle

\section{Introduction}
\label{sec:1} 

The tendency to favour information that is consistent with our 
beliefs --- this so-called confirmation bias is ``…perhaps the 
best known inferential error to come out of the literature on human reasoning'' (Evans 1989, p.41). It is ubiquitous in  human and animal cognition, but perhaps its most pernicious influence 
(superficially, at least) concerns human decision making. Confirmation bias 
has been linked to the spread of fake news (Stibel 2018), creation of echo 
chambers and polarisation (Flaxman \textit{et al}. 2016, Sikder \textit{et al}. 
2020, Starnini \textit{et al}. 2016), and the replication crisis in sciences 
(Lilienfeld 2017).  Confirmation 
biases have been demonstrated in diverse areas, 
from financial investment to medical diagnosis and legal proceedings 
(Croskerry 2003, Farmer 1999, Pang \textit{et al}. 2017, Roach 2010). 

The term confirmation bias is usually attributed to Wason (1960, 1961), who 
observed that naïve observers typically prefer to seek evidence with potential 
to confirm rather than falsify a given hypothesis. A confirmation bias appears 
to preclude rational belief, at least to the extent that rationality is understood 
as Bayesian inference. The rational status of Bayesian inference has been 
established both in human (e.g., Griffiths \textit{et al}. 2010) and in animal 
cognition (e.g., Ramirez \& Marshall 2017). Nevertheless, confirmation biases 
are not unique to humans (Odegaard \textit{et al}. 2018, Stolyarova \textit{et al}. 
2019), but in animal cognition, behavioural biases would translate to survival 
disadvantage. This observation suggests that, contrary to the prevailing 
dogma, there may be a powerful evolutionary advantage in confirmation 
bias, a postulate that we explore here. 

Section~\ref{sec:2} reviews arguments that suggest an 
adaptive interpretation of confirmation bias. Then, in Section~\ref{sec:3}, we 
define rational updating of prior beliefs when explaining observations or data. 
Our definition is based on the notion of biological efficiency (Brody \& 
Trewavas 2022), and leads to the familiar Bayesian updating in simple cases. 
In Section~\ref{sec:7}, we consider an active form of confirmation bias for 
evidence selection, relative to binary hypothesis testing. We extend the 
classical Bayesian analysis by considering a family of mutually inequivalent 
information sources that do not admit joint probability distributions. 
Confirmation biases are shown to arise when minimising the 
expected error probability in hypothesis selection. 

The analysis is extended in Sections~\ref{sec:8} and \ref{sec:9} to the case 
of a sequential information seeking. Here, we establish the remarkable fact 
that an optimal level of confirmation bias leads to both a minimum memory 
requirement and an exponential reduction in error probability. This is a powerful result, which reveals a surprising rationality for confirmation bias (under particular conditions). In Section~\ref{sec:10}, we link these 
results to the idea of active inference, whereby the decision maker seeks to 
optimise the expected information gain, and show in our example that the 
optimal strategy agrees with what follows from minimising expected error 
probability. 
The fact that we are exploring the space of inequivalent information sources 
means that our search has to be carried out not on a conventional probability 
space, but rather on a Hilbert space of square-root probabilities, using the 
machinery of quantum probability rules, in line with formalisms of quantum 
cognition (Pothos \& Busemeyer, 2022). Our finding of an exponential 
reduction in error probability provides the first example in which the so-called 
``quantum supremacy'' leads to a significant evolutionary advantage, as we 
explain in Section~\ref{sec:11}. On this view, our protocol provides the 
first operational way of implementing ``active quantum inference''. 
We conclude in Section~\ref{sec:12} with further discussion.

\section{Is a confirmation bias adaptive?}
\label{sec:2} 

Akin to a confirmation bias, it has been argued that a tendency to overestimate 
environmental correlations arises because of limitations in working memory. This 
behaviour can be adaptive, for, it helps with the early detection of 
associations in nature (Alloy \& Tabachnik 1984, Kareev 2000, Lopes 1982). A 
related idea concerns survival under scarce resources. In such cases, 
reinforcement learning analyses have shown that exploiting rare successes is 
beneficial, even at the expense of false positives (Ohta \textit{et al}. 2021, see 
also Farashahi \textit{et al}. 2019). More generally, learning differentially from 
positive compared to negative outcomes could be adaptive, depending on how 
they match reward structure (e.g., Daw \textit{et al}. 2002, Lefebvre \textit{et al}. 
2022). For example, with agent-based modelling, Bergerot \textit{et al}. (2024) 
suggests that a (moderate) confirmation bias can be adaptive in group decision 
making, depending on resource scarcity. However, human studies do not 
always reveal confirmation bias sensitivity to reward structure (Gershman 2015). 
More generally, such approaches cannot help explain the putative rational status 
of confirmation biases independently of reward structure, even if the mind can 
faithfully track it. 

Others have proposed that confirmation biases reflect bounded-rational Bayesian 
inference. For example, repetition of previous choices, regardless of evidence, 
would be a simpler action policy, whatever the evidence (Gershman 2020); or, an 
agent might find it more efficient to focus on an initial hypothesis (Doherty \& 
Mynatt 1986). Indeed, information overload has been linked to increased 
confirmation bias (Goette \textit{et al}. 2020). Another approach has been to 
explain confirmation biases based on the interaction between belief updating and 
source reliability, with an additional independence assumption between beliefs and 
source reliability (Pilgrim \textit{et al}. 2024). This is an interesting approach, though 
the independence assumption may be situational (e.g., Madsen \textit{et al}. 2020). 
Considering bounded-rational models of Bayesian inference in general, such as 
models whereby biases arise from noise (Costello \& Watts, 2014) or sampling 
problems (Zhu \textit{et al}., 2020), it is hard to identify explanations for confirmation 
biases. In summary, existing work provides limited insights regarding the potential 
rational status of confirmation biases.

\section{Rational updating of belief} 
\label{sec:3} 

Let us begin with the basic criterion on how 
information should be processed rationally for belief updating, 
because a confirmation bias is concerned with how we seek and 
interpret information. We consider the situation in which one obtains 
information about a matter of interest under uncertainty. We shall refer 
to such information as “evidence” or “observation”. A typical evidence 
will be inconclusive, but it nevertheless provides partial information 
about the subject of interest, based on which one updates 
their prior belief to a posterior belief.  

We wish to identify the condition such an updating ought to fulfil so as to 
\textit{define} what we mean by ``rational''. To this end, we note that any 
action a biological system might take --- in the service of information 
acquisition --- results in energy consumption, so that survival can be considered equivalent to 
minimising energy consumption. We are assuming here an 
equivalence between energy and information (Landauer 1961, Ororbia \& 
Friston 2023). Hence, the criterion we impose is that an update of belief 
must be such that it minimises averaged energy consumption resulting 
from the belief update, thus leading to biological efficiency.

This criterion is met if the posterior is the one that is the 
closest to the prior, amongst all posteriors that are consistent with the 
evidence (Brody \& Trewavas 2022). In Bayesian statistics, this can be 
measured with the Kullback-Leibler divergence between the posterior 
and prior. If the evidence obtained is 
conclusive, then there is only one consistent posterior,  
but otherwise there can be many that are consistent with an observation, 
providing equally accurate accounts of the data. Our criterion 
demands that amongst these Bayesian updated beliefs, the one that requires the 
smallest change relative to the prior must be chosen. This amounts to belief 
updating using only the arrival of new information --- called ``innovation'' in 
information science --- and this is ensured if the posterior is the one that is 
nearest to the prior (cf. Brody 2027). 

To formalise this condition, we need the notion of distance in the 
space of probabilities, and this is achieved by working not with probabilities 
but with square-root probabilities (Rao 1945). That is, if $\{p_k\}$ denotes the 
set of priors and $\{\pi_k\}$ the posteriors, then we consider the 
transformation $\sqrt{p_k}\to\sqrt{\pi_k}$. The space of square-root 
probabilities --- endowed with a Euclidean inner product --- is known as Hilbert space, in which 
each set of probabilities identifies a point on the unit sphere centred at the origin. 
The nearness of two probabilities can then be determined by the angle --- known 
in statistics as the Bhattacharyya distance 
(\textcolor{blue}{Supplemental Material A})  
--- between the two unit vectors of 
square-root probabilities (Brody \& Hook 2008). In what follows we shall be 
working with this angular separation between probabilities. 

In this spherical representation of beliefs (Brody 2023), given an observation, 
there are in general many points on the sphere that will be consistent with some 
evidence. A rational belief update $\sqrt{p_k}\to\sqrt{\pi_k}$ therefore is defined 
by the one for which its spherical distance to $\sqrt{p_k}$ is the smallest. Three 
important consequences follow (Brody \& Trewavas 2022). First, if we square the resulting posterior, 
we recover the Bayes rule for the probability conditional on the evidence 
(see \textcolor{blue}{Supplemental Material B} for a concrete example). This 
follows from a 
well-known result in probability that conditional probability is given by a 
projection on Hilbert space (Jacod \& Protter 2004). Second, the procedure 
for obtaining the transformation 
$\sqrt{p_k}\to\sqrt{\pi_k}$, satisfying our criterion, is known in the literature of 
quantum probability --- the mathematical framework that models behaviours 
of quantum systems --- as the ``von Neumann-L\"uders projection postulate''. 
The latter, however, is more general than the Bayes rule because --- in the 
context of a sequence of evidence gathering --- if successive observations do 
not admit a joint probability distribution, then the Bayes formula 
does not exist, whereas the projection scheme remains applicable. Third, and 
perhaps most importantly from a biological point of view, our transformation 
rule minimises, on average, the uncertainty about the matter of interest.  
Classically, under the free energy principle, this uncertainty can be read as 
the path integral of variational free energy, which 
provides a bound on self-information (Friston 2013). 

The rational belief updating defined here will be applied in 
the discussion below, whenever a decision maker obtains an external signal, 
irrespective of whether the signal is actively sought. If, however, there is a 
range of observations to choose from, then an additional criterion for rationality 
needs to be imposed, which will be next discussed

\section{An active form of confirmation bias}
\label{sec:7} 

Our main goal is to provide a rational explanation for an active form of 
confirmation bias. The setup we have in mind is whereby a person has a 
choice for seeking evidence, i.e. the observer can select among different 
sources of information. For example, we can think of a person who is 
concerned with how current affairs might impact future energy prices, and 
wishes to choose among different newspapers to gather evidence. The 
question then is, which newspaper is this person going to select.

An active form of confirmation bias is that people tend to choose an information 
source that aligns with their prior views. Intuitively, this is not surprising from 
the uncertainty-minimising feature of rational choice. That is, seeking information 
from a source that undermines one’s views will inevitably increase entropy, which 
is not desirable. However, an active form of confirmation bias cannot be fully 
explained using this argument alone. Instead, an adaptive and rational 
explanation of active confirmation bias can be obtained if we widen the scope of 
analysis by allowing for the possibility that random variables modelling different 
observations need not admit a joint probability density.

For concreteness, we consider the case of a binary selection. The objective  
is to find the best evidence that will minimise the expected probability of selecting 
the wrong hypothesis out of two possible hypotheses. Each evidence, or 
observation, yields two possible outcomes --- neither will be conclusive for 
choosing the correct hypothesis, but they provide partial information. An 
observation is usually modelled by a random variable, but in the Hilbert space 
setup it is represented by a matrix. A parametric family of observations giving 
binary outcomes can be modelled in a matrix form by 
\begin{eqnarray}
{\hat\xi}(\theta) = \left( \begin{array}{cc} \cos\theta & \sin\theta\\ 
\sin\theta & -\cos\theta \end{array} \right) \, .
\label{eq:z5}
\end{eqnarray}  
Then for each value of $\theta\in[0,\pi)$ the eigenvalues of ${\hat\xi}(\theta)$ 
are given by $\pm1$, representing the binary outcomes resulting from sampling 
the source of evidence. Following from the discussion above, $\theta$ could be 
seen as the degree of bias (for example, an editorial bias) in different newspapers, 
regarding whether energy prices are likely to increase or not. 
 
The parameter $\theta$ allows us to measure the distance between a given 
hypothesis and a given observation, which is a measure the degree of bias. 
How does this work? Recall that every hypothesis can be mapped to a random 
variable. Hypothesis $k$ thus asserts that random variable ${\hat\xi}(\theta_k)$, 
modelled as a matrix, takes the value 1. Then from the rational projection rule 
it follows that the probability of obtaining an affirmative answer from sampling 
the evidence ${\hat\xi}(\theta)$ when Hypothesis $k$ is the correct hypothesis 
is given by 
\begin{eqnarray}
b_k(\theta) = \cos^2 \half (\theta-\theta_k)  
\label{eq:z6}
\end{eqnarray}
for $k=1,2$. Note that (\ref{eq:z6}) is known as the ``Born probability rule'' 
in quantum theory, which is typically presented as a postulate and not a derived 
concept within the theory. However, here it can be shown that (\ref{eq:z6}) 
follows as a consequence of the rational belief updating principle defined in 
Section~\ref{sec:3}. In this setup, evidence selection corresponds to selecting 
a particular $\theta$, and we wish to find $\theta=\theta^*$ that minimises the 
probability of choosing the wrong hypothesis.

Although we consider a binary example here, the framework can be extended 
to multiple hypotheses and/or observations returning multiple possible outcomes. 
The important points in the present formulation --- stemming from the use of 
Hilbert space --- are (i) that we are able to deal with observations that need not 
admit joint probabilities, and (ii) that we are able to introduce the notion of a 
metric in the space of hypotheses (decisions) and inferences (observations). 
This latter point is important in analysing an active form of confirmation bias: we 
want to examine situations where people seek information sources that 
are more aligned, or close to, their prior views, but without the notion of a 
metric it is not possible to determine how close these are. In the present example, 
the distance between an information source ${\hat\xi}(\theta)$ and 
Hypothesis $k$ is given by $|\theta-\theta_k|$, and similarly the separation of 
the two hypotheses is given by $|\theta_2-\theta_1|$. 

\begin{figure}[t]
  \centering
       {\includegraphics[width=0.48\textwidth]{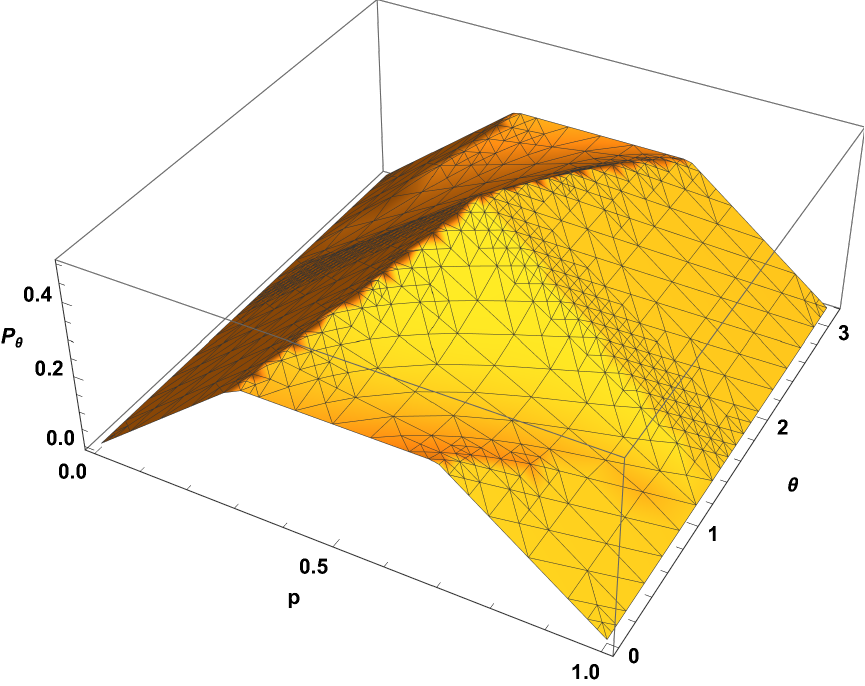}}\hfill
       {\includegraphics[width=0.48\textwidth]{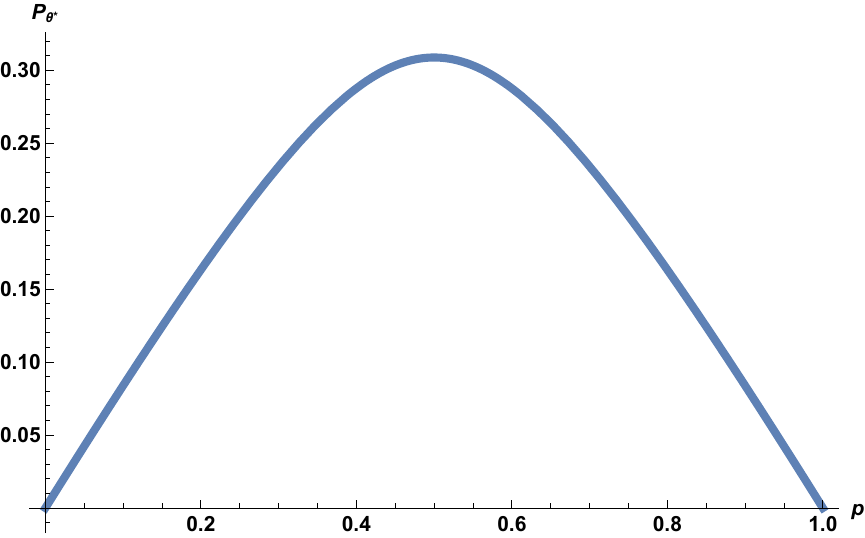}}\hfill
\caption{\textit{Error probability}. The error probability $P_\theta(p)$ is
plotted here on the left panel as a function of the prior $p\in[0,1]$ and 
the evidence choice $\theta \in [0,\pi)$. 
The parameters are set at $\theta_1=5\pi/8$ and $\theta_2=3\pi/8$. 
For each prior $p$ there is a value $\theta^*$ such that the error probability 
is minimised. For each $p$, the resulting error probability $P_{\theta^*}$ for 
the optimal strategy is shown in the right panel. 
} 
\label{fig:z1} 
\end{figure}

A rational decision maker seeks the evidence ${\hat\xi}(\theta^*)$ that 
minimises the (error) probability of selecting the incorrect 
hypothesis. If $p$ is the prior for Hypothesis 2, then the optimal inference 
strategy, if no evidence is obtained, is 
to choose Hypothesis 2 if $p>1/2$, and Hypothesis 1 if $p\leq1/2$. Hence 
the error probability as a function of prior is simply $P(p)=p$ for $p<1/2$ 
and $P(p)=1-p$ for $p\geq1/2$. However, once a piece of evidence has been 
selected, corresponding to some $\theta$, then depending on the outcome 
$\xi=\pm1$ the prior is 
updated to the posterior $\pi(\xi)$, according to the Bayes formula 
(see \textcolor{blue}{Supplemental Material D}). 
Hence, the optimal strategy is to choose Hypothesis 2 if $\pi(\xi)>1/2$, and 
Hypothesis 1 if $\pi(\xi)\leq1/2$, depending on the sampled outcome 
$\xi=\pm1$. However, for a 
fixed $\theta$ such that $b_2(\theta)>b_1(\theta)$, if the prior $p$ is 
larger than $(1-b_1)/(2-b_1-b_2)$, then $\pi(\xi)>1/2$ 
irrespective of the outcome $\xi$ (see \textcolor{blue}{Supplemental Material D}). 
In this case, Hypothesis 2 will always be 
selected, so that the error probability is $P_\theta(p)=1-p$. Similarly, if 
$p<b_1/(b_1+b_2)$ then Hypothesis 1 will always be selected, so that the 
error probability is $P_\theta(p)=p$. In the intermediate case, the error probability 
associated with the optimal strategy is given by 
\begin{eqnarray}
P_\theta(p) = p(1-b_2(\theta))+(1-p)b_1(\theta) \, , 
\label{eq:z8} 
\end{eqnarray} 
where $b_k(\theta)$ is given in (\ref{eq:z6}). In Figure~\ref{fig:z1} we sketch 
an example of the error probability as a function of the prior $0\leq p\leq1$ 
and the evidence choice $0\leq\theta<\pi$. 

A rational individual seeks to minimise, for a given prior $p$, the 
error probability, by choosing the optimal evidence ${\hat\xi}(\theta^*)$. This 
is found by differentiating $P_\theta(p)$ with respect to 
$\theta$ and solving for zero, which determines $\theta^*(p)$ 
(see \textcolor{blue}{Supplemental Material E}). 
The result is revealing because it identifies the separation between the 
optimal evidence choice $\theta^*(p)$ for a given level of prior $p$ and the 
two alternative choices $\theta_1$ and $\theta_2$. The question then is 
whether the information source one should seek becomes closer to the hypothesis, 
when the prior for that hypothesis increases.  

\begin{figure}[t]
  \centering
       {\includegraphics[width=0.48\textwidth]{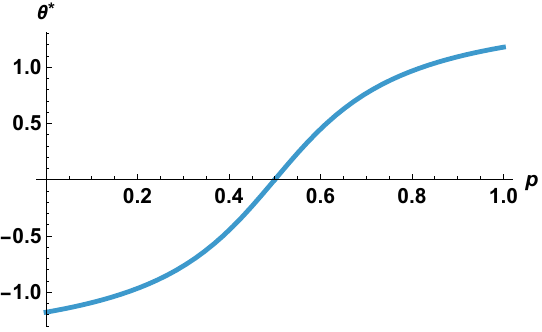}}\hfill
        {\includegraphics[width=0.48\textwidth]{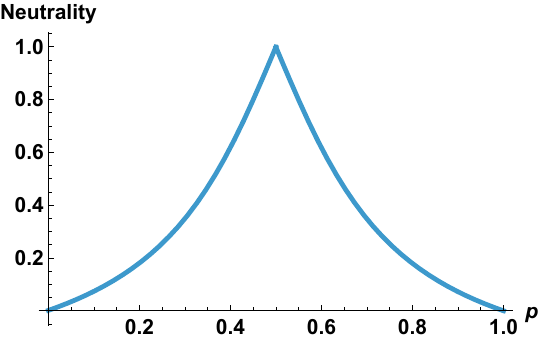}}\hfill
\caption{\textit{Optimal inference and confirmation bias}. The optimal inference 
choice $\theta^*(p)$ is plotted in the left panel as a function of the prior $p$, for the 
parameters $\theta_1=5\pi/8$ and $\theta_2=3\pi/8$. Note that $\theta^*$ and 
$\theta^*\pm n\pi$ for $n=0,1,2,\ldots$ represent identical pieces of evidence, so to avoid a 
discontinuity in $\theta^*(p)$, we have plotted $\theta^*(p)-\pi$ when $p<1/2$ 
and $\theta^*(p)$ when $p\geq1/2$. On the right panel the corresponding 
degree of neutrality 
$N(p)$ in evidence seeking is plotted as a function of the prior $p$. The neutrality 
is unit normalised such that the value $N(p)=1$ represents unbiased query, 
whereas $N(p)=0$ represents a 100\% biased search. 
} 
\label{fig:z2} 
\end{figure}

In Figure~\ref{fig:z2} (left panel) we plot the optimal inference choice 
$\theta^*(p)$ as a function of the prior $p$. We find that the the 
optimal choice $\theta^*$ is unbiased when $p=1/2$, whereas 
$\theta^*\to\theta_2$ as $p\to1$, and $\theta^*\to\theta_1$ 
($\!\!\!\!\mod \pi)$ as $p\to0$, in a monotonic manner. In other words, the 
information source is biased towards where the prior view is 
stronger. We can define the normalised degree of neutrality 
$N(p)$ by the separation distance $|\theta^*(p)-\theta_2|$, normalised by 
its maximum value over all priors, for $p>1/2$, and by 
$|\theta^*(p)-\theta_1|$ for $p\leq1/2$. 
An example of the neutrality measure $N(p)$ is also sketched in Figure~\ref{fig:z2} 
(right panel). We see that for $p=1/2$ the evidence choice is unbiased, 
given by $\theta^*(0.5)=(\theta_1+\theta_2-\pi)/2$, whereas in the limits 
$p\to0$ and $p\to1$ the evidence choice is fully biased.

Summing up so far, the key conclusion is that the active form of confirmation bias revealed here 
is a result of a rational decision-making, based on the criterion of minimising expected 
error probability. The neutrality measure $N(p)$ can be viewed as a 
consequence of rational behaviour. However, it is possible to identify even stronger results, concerning the rational essence of the confirmation bias, when we consider a sequential 
information gathering, which we do next.

\section{Sequential evidence search and memory capacity}
\label{sec:8} 

We considered the case in which evidence is sought just once 
and identified the optimal information source that minimises the error probability 
associated with hypothesis selection. More commonly, an agent will be seeking 
information successively. According to the optimality criterion of minimising 
expected error probability, evidence should be initially selected based on the prior.  Once evidence is sampled, the prior is updated to the posterior, which now 
becomes the prior for selecting the next observation. We are interested in how 
this updating works.

To this end, we first consider the posterior probability $\pi(\xi)$ under the 
optimal inference choice $\theta=\theta^*$. Substituting the expression for 
$\theta^*(p)$ in 
(\ref{eq:z6}) and then using the result in the Bayes rule for $\pi(\xi)$ we deduce, 
after some algebra (see \textcolor{blue}{Supplemental Material E}), that 
\begin{eqnarray}
\pi^*(\xi) = \frac{1}{2}\left( 1+\xi\sqrt{1-4p(1-p)\cos^2\half\delta}\right) ,  
\label{eq:z10} 
\end{eqnarray}
where we have defined $\delta=\theta_2-\theta_1$; recall that $\xi=\pm1$ 
denotes the outcome from sampling the first piece of evidence. 
In an optimal sequential evidence gathering situation, the posterior 
$\pi^*(\xi)$ for an initial outcome $\xi$ 
now becomes the prior for the second evidence sampling. This means 
that by substituting the expression for $\pi^*(\xi)$ in $p$ in the right side of 
(\ref{eq:z10}), with $\xi$ now the outcome of the second evidence sampling, we 
obtain the posterior after the second evidence has been sampled. 
The remarkable feature of the optimal posterior is the following identity: 
\begin{eqnarray}
\pi^*(\xi) \left(1-\pi^*(\xi) \right) = p(1-p) \cos^2\half\delta \, . 
\label{eq:z11} 
\end{eqnarray} 
That is, the expression for $\pi(1-\pi)$ is independent of the outcome $\xi$ of 
the first evidence. It follows 
that the posterior probability $\pi_n^*$ after sampling $n$ pieces of evidence optimally, 
with an appropriate level of confirmation bias, is given by 
\begin{eqnarray}
\pi_n^*(\xi) = \frac{1}{2}\left( 1+\xi\sqrt{1-4p(1-p)\cos^{2n}\half\delta}\right) ,  
\label{eq:z12} 
\end{eqnarray}
where $\xi$ now represents the sampled outcome of the final evidence. (An 
inductive proof of (\ref{eq:z12}) was first obtained in Brody \& Meister 1996a, 
although the proof here is considerably simpler.) 

To understand the significance of (\ref{eq:z12}) let us compare this to the 
case where the observable ${\hat\xi}(\theta)$ is fixed and there is no flexibility 
of choosing evidence to sample, thus confirmation bias is prohibited.   
In this case, after $n$ sampling there are $2^n$ possible 
``histories'' for the sampled evidence. However, on account of the 
Markovianity, the posterior probability after $n$ trials only depends 
on the total number of positive outcomes in the sample 
(see \textcolor{blue}{Supplemental Material D}), and not on their order. 
Hence, in terms of the memory 
required for hypothesis testing, an 
optimal procedure provides a reduction $2^n\to n$ from exponential to 
linear growth. 

In contrast, if an active form of confirmation bias is permissible, and if 
evidence is gathered optimally, then we have a remarkable reduction 
$2^n\to 2$. That is, we only need to know the outcome of the final trial! 
In other words, 
an appropriate level of confirmation bias allows the decision maker to `forget' 
about the past history except for the most recent event, thus getting away with 
the smallest amount of memory capacity for inference. Intuitively, if we are allowed to select 
evidence in a way that depends upon accumulated posterior beliefs, there is no 
need to remember everything we have seen in the past: a Bayes-optimal 
inference can be made purely on the basis of the current observation, because 
that observation is informed by our past experience. Evidently, this remarkable 
feature offers a significant evolutionary advantage and speaks to the importance 
of selecting pieces of evidence using a confirmation bias.

\section{Exponential reduction in the error probability}
\label{sec:9} 

Another aspect of confirmation bias that interests us here is how the error 
probability reduces with sample size, when compared to the case 
in which only one ``unbiased'' evidence is available (i.e. there is no 
opportunity for active selection). For this purpose, we follow the same 
technical manipulations as in (\ref{eq:z10}). That is, we substitute the 
expression for $\theta^*(p)$ in (\ref{eq:z6}) and then in (\ref{eq:z8}). Then, 
after some algebra, we deduce that the error probability after a single evidence 
sampling is 
$P_{\theta^*}(p) = \frac{1}{2}( 1-\sqrt{1-4p(1-p)\cos^2\half\delta})$, 
with $\delta=\theta_2-\theta_1$. This result can 
alternatively be deduced from the fact that, if the evidence is chosen optimally, 
then it remains true from the optimal hypothesis selection rule that the error 
probability is $P_{\theta^*}(p) = \pi^*(\xi)$ if $\xi=-1$ and $P_{\theta^*}(p) = 
1-\pi^*(\xi)$ if $\xi=+1$ --- in both cases we deduce this result from 
(\ref{eq:z10}). 

\begin{figure}[t]
  \centering
  {\includegraphics[width=0.480\textwidth]{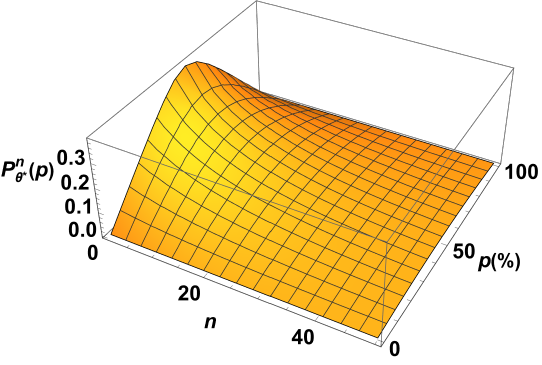}}\hfill 
       {\includegraphics[width=0.480\textwidth]{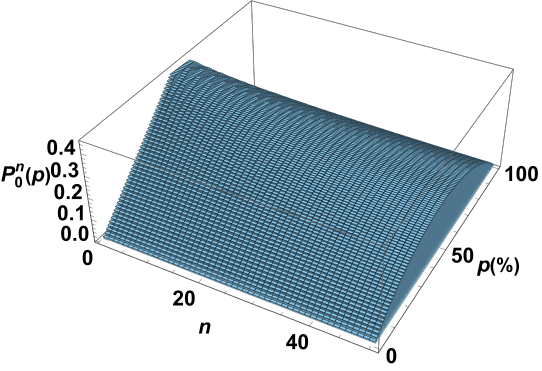}}\hfill
\caption{\textit{Error probabilities with and without confirmation bias}. The 
error probability $P_{\theta^*}^{(n)}(p)$ is plotted here in the left panel as a 
function of the prior $p$ and the sample size $n$. For each fixed $p$ the 
error probability decreases exponentially in $n$. The result is compared in 
the right panel to the (suboptimal) 
case where there is no freedom in choosing the evidence, that is, each 
evidence is modelled by ${\hat\xi}(0)$. The result shows that the error 
probability decreases very slowly. The parameters are chosen here such 
that $b_1(0)=0.47$ and $b_2(0)=0.58$. 
} 
\label{fig:z3} 
\end{figure}

As in the case of the posterior analysis in sequential sampling, if the prior 
$p$ is updated to the posterior $\pi^*(\xi)$, then by substituting this in place 
of $p$ in $P_{\theta^*}(p)$, we obtain the error probability conditional on the 
outcome $\xi$ of the first evidence sampling. To obtain the error probability 
we average the result over $\xi$. However, from 
(\ref{eq:z11}), we see that the updated error probability $P_{\theta^*}(\pi^*(\xi))$ 
is independent of $\xi$. It 
also follows that by iterating the process $n$ times, we find the expression 
for the error probability after optimally sampling $n$ pieces of evidence: 
\begin{eqnarray}
P_{\theta^*}^{(n)}(p) = \frac{1}{2}\left( 1-\sqrt{1-4p(1-p)\cos^{2n}\half\delta}\right) 
\sim p(1-p) \exp\left( \ln\left( \cos^{2}\half \delta \right) \, n \right) \, .
\label{eq:z14} 
\end{eqnarray}
The second expression on the right shows the asymptotic behaviour 
for large $n$.  
Hence, the error probability in identifying the correct hypothesis reduces exponentially fast, when samples are 
chosen optimally. This is in contrast to the case in which there is only a 
single source of evidence and without a confirmation bias. In this latter case, the 
analysis of the error probability, although elementary, is a little cumbersome 
(see \textcolor{blue}{Supplemental Material D}): the result shows that the 
error probability decays very slowly. In Figure~\ref{fig:z3}, we compare the 
dependence of the error probabilities on sample size with (optimal) 
confirmation bias and without confirmation bias. In sum, we find that an 
appropriate level of confirmation bias, in addition to reducing demands on 
memory capacity, leads to an exponential reduction in error probability. 
This illustrates another key way in which a confirmation bias has clear evolutionary value.

\section{Confirmation bias and active inference}
\label{sec:10} 

We have examined optimal sampling or search from the perspective of 
minimising expected error probability associated with hypothesis selection. 
A complementary approach is to maximise the expected information gain 
(i.e. mutual information) following each sample. This consideration naturally 
takes us to the realm of ``active inference'' (Parr \textit{et al}. 2022). In active 
inference, a decision maker chooses the evidence that minimises expected 
free energy. This amounts to maximising the mutual information between 
observations and their inferred causes. 

Technically, expected free energy is comprised of expected information gain 
and expected cost, where cost scores the violation of prior beliefs about 
outcomes that are characteristic of the decision maker in question (Da Costa 
\textit{et al}. 2020). From a statistical perspective, these two components 
render decision-making Bayes optimal, in the dual 
sense of optimal experimental design and Bayesian decision theory, 
respectively (Lindley 1956, Berger 2011). 

In the present context, however, the classical framework of active inference 
is not directly applicable, on account of the fact that the two random variables 
represented by the matrices ${\hat\xi}(\theta)$ and ${\hat\xi}(\theta')$, for 
$\theta\neq\theta' ~(\!\!\!\!\mod \pi)$, do not admit a joint probability distribution. 
Practically, what this means is that if evidence ${\hat\xi}(\theta')$ is sampled 
after ${\hat\xi}(\theta)$, then the resulting posterior will not agree with the 
two pieces of evidence sampled in reverse order: a situation that is commonly observed in human 
cognition (Busemeyer \& Wang 2015, Pothos \& Busemeyer 2013, 2022, 
Haven \& Khrennikov 2016, Trueblood \textit{et al}. 2017). 
Hence, we must formulate active inference on a Hilbert space. 
The problem can be re-stated as follows. A 
decision maker is presented with a set of observations ${\hat\xi}(\theta)$ to 
choose from, to make an inference about the two hypotheses. The task is to 
identify the observation that provides the most information about the correct 
hypothesis. 

Information here is measured with mutual information, 
which is the relative entropy between the joint 
density of evidence $\theta$ and the hypothesis, and the 
product of their marginal densities. Note that while different observations do 
not admit a joint density, for every observation the joint density of the 
observation and the hypothesis does exist, making mutual information 
well defined (this joint density is known as a ``generative model'' in active 
inference). The result defines the amount of information contained in the 
evidence about the correct hypothesis.

\begin{figure}[t]
  \centering
  {\includegraphics[width=0.480\textwidth]{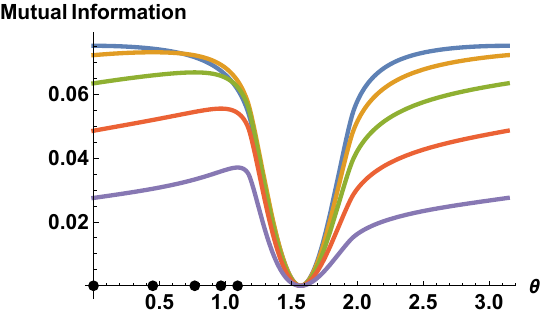}}\hfill 
  {\includegraphics[width=0.480\textwidth]{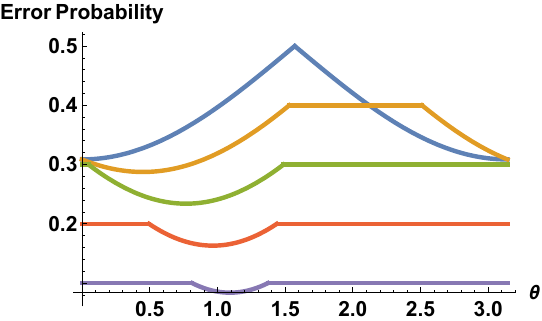}}\hfill 
\caption{\textit{Mutual information and error probability}. The mutual 
information associated with the evidence ${\hat\xi}(\theta)$ is plotted here 
on the left panel as a function of $\theta\in[0,\pi)$ for the priors 
$p=0.5,0.6,0.7,0.8,0.9$ (from blue to purple). 
The corresponding values of $\theta^*(p)$ are 
shown by the dots on the horizontal axis. It is evident, for each $p$, that 
the mutual information is maximised at the value $\theta=\theta^*(p)$. 
The parameters are set at $\theta_2=3\pi/8$ and $\theta_1=5\pi/8$. 
The corresponding error probabilities $P_\theta(p)$, the minimum of 
which for each $p$ determines the value of $\theta^*(p)$, for the same 
$p$-values, are shown on the right panel for comparison. 
} 
\label{fig:z4} 
\end{figure}

With this in mind, in Figure~\ref{fig:z4} we plot the mutual information as a 
function of the evidence choice $\theta$ for various values of the prior $p$. 
The results confirm that the mutual information is indeed maximised for each 
$p$ at the optimum $\theta=\theta^*(p)$ --- thus, the confirmation bias is 
recovered. In general, optimisation 
based on the error probability is not equivalent to optimisation based on mutual 
information. Nevertheless, here we find that the optimal piece of evidence 
$\theta^*(p)$ that minimises the error probability does indeed maximise the 
mutual information. It therefore 
follows that confirmation bias is a rational and adaptive phenomenon from the 
point of view of active inference, when the latter is extended to a Hilbert space 
and random variables are represented by noncommuting matrices. 

The error probability $P_\theta(p)$ is also shown in Figure~\ref{fig:z4} for 
comparison. The two criteria --- maximisation of the mutual information 
and minimisation of the error probability --- both lead to the same optimal 
sampling strategy, which reflects a confirmation bias. Note, though, that they 
need not agree when it comes to suboptimal 
strategies due to other constraints. We have assumed that pieces of evidence 
${\hat\xi}(\theta)$ for all $\theta\in[0,\pi)$ are available without constraints, but 
in reality, often only finitely many information sources are available. 
For example, if the sample in question concerns the choice of newspaper, then 
perhaps there are only a dozen or so available.  

To illustrate, consider the example shown in Figure~\ref{fig:z4}, with  
$p=0.7$. In this case, the optimal observation is $\theta^*\approx0.768$, but 
suppose that there are only two observations available, corresponding to 
$\theta=1.3$ and $\theta=2.5$ (in the newspapers example, there would be 
only two newspapers). In this case, if the criterion is to minimise the 
error probability, then we find $P_{1.3}=0.27$ and $P_{2.5}=0.3$, so the best 
available sample is given by $\theta=1.3$. On the other hand, if the criterion 
is to maximise the amount of information gain, then we have $J_{1.3}=0.031$ 
and $J_{2.5}=0.057$, so the best available sample now is given by 
$\theta=2.5$. Therefore, if sampling of evidence is constrained, it is an open 
question whether people generally follow the maximum information 
criterion or the minimum error probability criterion, or perhaps yet another 
criterion.

\section{Relation to quantum hypothesis testing}
\label{sec:11} 

We have shown that the empirically observed phenomenon of 
confirmation bias emerges as a consequence of an adaptive and 
rational behaviour when formulated on the 
Hilbert space of square-root probabilities. The key ingredient 
is the introduction of a range of inequivalent 
information sources --- two observations are said to be inequivalent 
if their matrix representations on 
Hilbert space are not commutative, or, putting it differently, their 
representations as random variables on a probability space do not admit a 
joint probability distribution. 

A formulation of modelling random phenomena on Hilbert space, in itself, 
is an old idea in statistics, known as ``information geometry'' 
(Brody \& Hook 2008). 
When endowed with the introduction of incompatible 
random variables, we are in the realm of ``quantum probability''. Confirmation 
bias can therefore be viewed as an adaptive behaviour, when we extend 
statistical analysis formulated on Kolmogorovian axioms into the quantum 
domain. 
Indeed, over the past two decades it has been shown that many of 
seemingly paradoxical human behaviours can be explained using models 
based on  quantum probability theory, leading 
to the creation of the emerging research field of 
``quantum cognition'' (Pothos \& Busemeyer 2022). 
Our contribution brings in the tool of quantum hypothesis testing in this 
endeavour of trying to understand human behaviour as optimal search.
Note, our formulation can be cast within the general theory of quantum 
hypothesis testing (Helstrom 1967, Holevo 1973, Stratonovic 1973, Yuen 
\textit{et al}. 1975, Belavkin 1975, Helstrom 1976) --- see 
\textcolor{blue}{Supplemental Material E}. 
The combination of quantum hypothesis testing and active inference, on the 
other hand, leads to the notion of ``active quantum inference''. The idea that 
active inference --- and the free-energy principle that underpins it --- can be 
extended into the quantum domain has been envisaged previously 
(Fields \textit{et al}. 2022). 
The current formulation shows that the maximisation of mutual information 
over the space of noncommutative observations provides 
the first operationally meaningful --- and conceptually straightforward --- way 
of extending active inference into the quantum domain. 

When we implement the analysis on Hilbert space, a key ingredient that we 
exploit implicitly is the concept of superpositions of states. This idea 
is often discussed in the context of quantum computing that leads to the so-called 
``quantum supremacy'' --- for example, in the 
problems of prime number factorisation and 
database search. A less-known example of quantum supremacy lies in the 
problem of sequential hypothesis testing, where the error 
probability is exponentially suppressed when compared 
to classical counterparts (Brody \& Meister 1996a, 1996b). 
Importantly, the present analysis --- concerning quantum supremacy in 
hypothesis testing --- does not require quantum resources in the physical 
sense. We merely require observations not admitting a joint probability distribution. 
Behaviourally, such incompatibility can simply reflect an inability to accurately 
assess two propositions simultaneously (perhaps because one invokes perspectives which alter our understanding of the other). Our findings here provide the first example within 
quantum cognition that proposes that human brain can exploit 
quantum supremacy (see \textcolor{blue}{Supplemental 
Material F and G} for further discussion on superpositions of state).

\section{Discussion}
\label{sec:12} 

In summary, we have considered an active form of confirmation bias --- 
concerning how people search for, or gather information --- by formulating 
the problem in the 
context of hypothesis testing on a Hilbert space. By imposing the criterion of 
minimum error probability, we have shown that the choice for optimal evidence 
indeed leads to an active confirmation bias.
In our analysis 
we have assumed the existence of incompatible observations, in the sense 
that the characterisation of different kinds of observations need not admit a joint 
probability distribution. This takes us to the domain of quantum probability, 
which seems inevitable if we are to model inequivalent information sources. 
We then considered an alternative criterion of information maximisation for 
sampling, and found that the optimal strategy led to the same confirmation 
bias. Hence, we  can read our approach as providing an operationally tractable scheme for an 
active quantum inference, thus opening up a range of opportunities for 
applications. 

The fact that the optimal strategy, invoking an appropriate level of confirmation 
bias, leads to both an exponential reduction in expected error probability and a 
minimum memory capacity is clearly significant in terms of evolutionary advantage. 
However, we have to be cautious about the effect of confirmation bias in society. 
It is worth noting that one assumption in our analysis is that every information 
source is genuine; meaning that it comprises a signal for the correct hypothesis 
and unbiased noise. We are implicitly assuming that available observations are not 
designed to be deceptive. This assumption is not unreasonable when applied to a 
problem in physical science, but in social science this assumption is not always valid. 
If an observation secretly contains an element that deliberately tries to misguide the 
decision maker (Brody 2022, Brody2026), then confirmation bias can be 
problematic for obvious reasons. This, of course, is outside of the scope of the 
present analysis, but it is worth keeping in mind that the apparent biological 
advantages of confirmation bias can be exploited to distort belief updating in 
arbitrary directions.

\vskip6pt

\vspace{0.3cm}
\begin{footnotesize}
\noindent {\bf Acknowledgements}. The author thanks Jim Al-Khalili, Eva-Maria Graefe and Simon Saunders  
for stimulating discussion. DCB acknowledges support from EPSRC 
(grant EP/X019926/1). EMP acknowledges support from EOARD grant FA8655-23-1-7220. KF is supported by funding from the Wellcome Trust (Ref: 226793/Z/22/Z).
\end{footnotesize}

\appendix

\section*{A. Rational updating of belief} 
\label{sec:1} 

Our definition of a rational updating of belief under uncertainty, namely, that 
an update of the prior must be such that the posterior is the one that is the 
closest to the prior, among all posteriors that are consistent with the evidence, 
can be illustrated using simple examples. To this end, suppose that the matter of 
interest is modelled by a random variable $X$ defined on a probability space, 
where $X$ takes on the value $x_k$ with the probability $p_k$ for $k=1,2,\ldots,N$. 
For instance, if one is interested in the weather in two days, because there is an 
event taking place, then $X$ may take four values $\{x_1,x_2,x_3,x_4\}$ labelling 
sunny, cloudy, rainy and snowy weathers. The numerical values assigned to the 
set $\{x_1,x_2,\ldots,x_N\}$ the label outcomes of chance can be significant in some 
applications to model cognitive behaviours (Brody 2026), but they do not concern 
us in the present discussion. 

The set of probabilities $\{p_k\}$ represents the \textit{a priori} view of the decision 
maker on the matter. An acquisition of evidence can then be modelled, for example, 
in the form of the detection of the random variable $\xi=X+\epsilon$, where $\epsilon$ 
represents noise, a random variable that is independent of $X$, that obscures the 
evidence. Alternatively the evidence may take the form $\xi=X\epsilon$, or yet 
another form. For example, if $X$ represents the weather in two days, then looking 
into a weather forecast --- an ``evidence'' or an ``observation'' --- provides partial 
information about the 
value of $X$, i.e. about the weather in two days. From the modelling perspective, 
the important point here is that we have two unknowns, $X$ and $\epsilon$, but only 
one known, $\xi$, from which the correct value of $X$ cannot be determined in 
general. Nevertheless, the value of $\xi$ provides partial information about the 
correct inference for $X$, based on which the prior view $\{p_k\}$ is updated to the 
posterior view $\{\pi_k\}$. 

The criterion of minimising overreaction asserts that this updating must be such 
that the set $\{\pi_k\}$ is the closest to the set $\{p_k\}$ without contradicting the 
evidence. Imagine that in the middle of the summer the weather forecast states 
that the temperature will drop significantly in two days time. In this scenario, an 
assessment that there is a high likelihood of a snow in two days for sure if 
consistent with the evidence (that the temperature will drop significantly), but it 
will be quite remote from the prior view of the weather in two days that already 
takes into account the fact that it is a midsummer day. The point is that there are 
many assessments that will be consistent with the evidence, but we are looking 
for one that is the closest to the previous assessment. 

To specify the notion of closeness among probabilities, it will be useful to work 
with square-root probabilities (Rao 1945). That is, we consider the transformation 
$\sqrt{p_k}\to\sqrt{\pi_k}$. It is, of course, certainly legitimate to consider the 
Euclidean distance of probabilities on a probability simplex, but the problem here 
is that distances get distorted on a probability simplex, especially near the 
boundaries, so that the separation of different beliefs cannot be measured fairly, 
and this motivates us to consider the square-roots instead. Now the space of 
square-root probabilities endowed with a Euclidean inner product is known as 
Hilbert space, in which each set of probabilities identifies a point on the unit 
sphere centred at the origin, because the sum of the squares of the square-root 
probabilities add up to one: 
\begin{eqnarray}
\sum_{k=1}^N \left( \sqrt{p_k}\right)^2 = 1 \, . 
\end{eqnarray}
The nearness of two probabilities can then be determined by the angle between 
the two unit vectors of square-root probabilities (Brody \& Hook 2008). 

One way of visualising this setup is to take the vectors $\vec{e}_1=
(1,0,0,\ldots,0)$, $\vec{e}_2=(0,1,0,\ldots,0)$, $\vec{e}_3=(0,0,1,\ldots,0)$,
and so on, of the states of no uncertainty, as the orthonormal basis elements of 
the Hilbert space. Then a generic vector 
\begin{eqnarray}
\left( \begin{array}{c} \sqrt{p_1} \\ \sqrt{p_2} \\ \vdots \\ \sqrt{p_N} \end{array} \right) 
= \sum_{k=1}^N \sqrt{p_k} \, \vec{e}_k 
\end{eqnarray}
can be seen to lie on the positive ``orthant'' of the unit sphere. If 
there is a second such vector $(\sqrt{\pi_1}, \sqrt{\pi_2}, \sqrt{\pi_3},\ldots,\sqrt{\pi_N})$, 
then their inner product can be equated to $\cos\theta$. Then 
\begin{eqnarray}
\theta = \cos^{-1} \left( \sum_{k=1}^N \sqrt{p_k \, \pi_k} \right) 
\end{eqnarray} 
defines the distance between the two probabilities $\{p_k\}$ and $\{\pi_k\}$. With 
the metric defined in this way, we need to identify the ``rational'' transformation 
$\{\sqrt{p_k}\} \to \{\sqrt{\pi_k}\}$ that minimises the distance $\theta$ of all 
$\{\sqrt{\pi_k}\}$ that are consistent with the evidence. We now work out such 
a transformation through an example.

\section*{B. Projection leads to the Bayes formula} 
\label{sec:5} 

In a Hilbert-space formulation, random variables are represented by symmetric 
matrices, whose eigenvalues correspond to the values they can take. A random 
variable $\xi$ that represents an evidence is then represented by a matrix 
${\hat\xi}$, whose eigenvalues represent the sampled values of $\xi$. The fact 
that an evidence is inconclusive means that the matrix ${\hat\xi}$ is degenerate, 
that is, it has degenerate eigenvalues. To explain this, let us consider the example 
$\xi=X+\epsilon$ for the evidence. This representation of an evidence is rather 
generic in that it takes the form of a familiar ``signal plus noise'' structure. 
Suppose that the ``noise'' 
random variable $\epsilon$ takes the value $\epsilon_m$ with the probability 
$q_m$. Then the eigenvalues of the matrix representation ${\hat\xi}$ of the 
random variable $\xi$ take the form $x_k+\epsilon_m$ for all index pairs 
$(k,m)$. It is entirely possible, however, to have a degenerate situation in 
which, for instance, 
\begin{eqnarray}
\xi_1=x_1+\epsilon_1, \quad 
\xi_2=x_1+\epsilon_2=x_2+\epsilon_1 , \quad \ldots  
\end{eqnarray}
holds. In this case, 
the detection $\xi=\xi_2$ of the evidence implies that the value of $X$ is 
either $x_1$ (and hence $\epsilon=\epsilon_2$) with some probability 
$\lambda$, or $X=x_2$ (and hence $\epsilon=\epsilon_1$) with some 
probability $1-\lambda$. Any choice of $\lambda\in(0,1)$ will be consistent with 
the evidence $\xi=\xi_2$, but the criterion of minimum overreaction asserts that 
the value of $\lambda$ must be chosen such that the \textit{a posteriori} 
probability is the closest to the \textit{a priori} probability. 

Now the \textit{a priori} probability that the random variable pair $(X,\epsilon)$ 
taking the values $(x_k,\epsilon_m)$, on account of their statistical independence, 
is given by the product $p_k \, q_m$. Hence the component of the Hilbert-space 
vector corresponding to the event $(X,\epsilon)=(x_k,\epsilon_m)$
is given by $\sqrt{p_k\, q_m}$. That is, in the large (tensor-product) Hilbert space 
representing all joint square-root probabilities of the random variable pair $(X, 
\epsilon)$, the prior state in a canonical basis is given by the vector $\sqrt{p_k\, q_m}$ 
for all index pairs $(k,m)$. If an evidence $\xi=\xi_n$ is detected, 
then all possible \textit{a posteriori} square-root probabilities must lie on the 
subspace (the eigenspace) of the Hilbert space associated with the eigenvalue 
$\xi_n$ of the matrix ${\hat\xi}$. In the above example of $\xi=\xi_2$ this will be 
the two-dimensional subspace spanned by the elements $\sqrt{p_1\, q_2}$ and 
$\sqrt{p_2\, q_1}$. In this example, an arbitrary transformation of the prior into a 
posterior that respects the observation thus takes the form 
\begin{eqnarray}
\left( \begin{array}{c} 
\sqrt{p_1q_1} \\ \sqrt{p_1q_2} \\ \vdots \\ \sqrt{p_1q_M} \\ \sqrt{p_2q_1} \\ 
\sqrt{p_2q_2} \\ \vdots \\ \vdots \\ \sqrt{p_Nq_1} \\ \sqrt{p_Nq_2} \\ \vdots \\ 
\sqrt{p_Nq_M} \end{array} \right) \Longrightarrow 
\left( \begin{array}{c} 
0 \\ \lambda \\ \vdots \\ 0 \\ 1-\lambda \\ 0 \\ \vdots \\ \vdots \\ 0 \\ 0 \\ \vdots \\ 
0 \end{array} \right) \, , 
\label{eq:S5} 
\end{eqnarray}
where $M$ is the number of values $\epsilon$ can take. Clearly, for all values of 
$\lambda\in[0,1]$ the right side of (\ref{eq:S5}) lie on a two-dimensional subspace 
of the Hilbert space. Our criteria therefore is to choose, of all those possibilities, 
the value of $\lambda$ such that the right side of (\ref{eq:S5}) is the closest to the 
left side. But because Hilbert space is endowed with a 
Euclidean inner product, it should be evident that the vector in this subspace 
that is the closest to the prior vector $\sqrt{p_k\, q_m}$ is the orthogonal projection 
of $\sqrt{p_k\, q_m}$ onto this subspace: 
\begin{eqnarray}
\left( \begin{array}{c} 
\sqrt{p_1q_1} \\ \sqrt{p_1q_2} \\ \vdots \\ \sqrt{p_1q_M} \\ \sqrt{p_2q_1} \\ 
\sqrt{p_2q_2} \\ \vdots \\ \vdots \\ \sqrt{p_Nq_1} \\ \sqrt{p_Nq_2} \\ \vdots \\ 
\sqrt{p_Nq_M} \end{array} \right) \Longrightarrow \frac{1}{\sqrt{p_1q_2+p_2q_1}}
\left( \begin{array}{c} 
0 \\ \sqrt{p_1q_2} \\ \vdots \\ 0 \\ \sqrt{p_2q_1} \\ 0 \\ \vdots \\ \vdots \\ 0 \\ 0 \\ 
\vdots \\ 0 \end{array} \right) \, .
\label{eq:S6} 
\end{eqnarray}

More generally, the relevant projector onto the subspace is given by 
${\mathds 1}\{\xi_n=x_k+\epsilon_m\}$, where ${\mathds 1}\{A\}$ denotes 
the indicator function for the event $A$, that is, ${\mathds 1}\{A\}=1$ if $A$ is 
true, and ${\mathds 1}\{A\}=0$ if $A$ is false. 
To identify the correct transformation rule consistent with our criterion of least 
overreaction, we note that a projection in Hilbert space does not preserve the 
length (the total probability) of the vector, so after an application of the projector 
${\mathds 1}\{\xi_n=x_k+\epsilon_m\}$ we must renormalise the vector to have 
unit length. With this in mind, we find that the correct transformation rule consistent 
with our requirement of least energy consumption is given by 
\begin{eqnarray}
\sqrt{p_k \, q_m} \to \sqrt{
\frac{p_k \, q_m}
{\sum_k p_k \, q_{n_k}}}
\, {\mathds 1}\{\xi_n=x_k+\epsilon_m\}\, ,
\label{eq:z2} 
\end{eqnarray}
where $n_k$ is the value of $m$ such that $\xi_n=x_k+\epsilon_m$, with the 
convention that $q_{n_k}=0$ if $n_k$ does not exist. 
Squaring the right-side of (\ref{eq:z2}) we find that this is nothing but the Bayes 
formula for the conditional probability ${\mathbb P}(X=x_k, \, \epsilon=\epsilon_m 
| \xi=\xi_n)$. In other words, our criterion implies the Bayes formula in simple 
circumstances, but our projection scheme is more general 
(see Ozawa \& Khrennikov 2023, Brody 2023). The generality follows 
from the fact that, suppose that after the detection of the evidence $\xi=X+\epsilon$ 
a second evidence of the form $\eta=Y+\epsilon'$ is detected, such that the random 
variable pair $(X,Y)$ does not admit a joint probability. In this case, the projection 
method remains applicable, whereas the Bayes formula does not exist. 

An important consequence of our criterion is that this transformation rule from the 
prior to the posterior minimises, on average, the uncertainty about the matter of 
interest (Brody \& Trewavas 2022). That is, of all posterior probabilities $\pi_k$ for 
the random variable $X$ that is consistent with the evidence $\xi=\xi_n$, the one 
given by 
\begin{eqnarray} 
\pi_k = \frac{p_k \, q_{n_k} }
{\sum_k p_k \, q_{n_k}} \, , 
\label{eq:z3} 
\end{eqnarray} 
which is just the Bayes formula for the conditional probability ${\mathbb P}
(X=x_k | \xi=\xi_n)$, minimises the uncertainty about $X$. In fact, we could have 
imposed the criterion of uncertainty minimisation as our guiding principle, and 
we would have arrived at the same conclusion. 

The uncertainty here can be measured either in terms of the variance, or by the 
Shannon-Wiener entropy, associated with the random variable $X$, conditional 
on the evidence $\xi=\xi_n$. Evolutional advantage of the criterion of uncertainty 
minimisation can be seen from the observation that entropy minimisation leads 
to the maximisation of free energy, which determines the amount of useful work 
that can be extracted from the environment of the inference maker. In general, if 
we view the purpose of a biological system as converting energy from the 
environment into useful work, then this criterion certainly is reasonable.

\section*{C. Confirmation bias for interpreting information}
\label{sec:6} 

The fact that a rational (Bayesian) updating minimises uncertainty carries a 
significant implication in understanding certain human behaviours. Suppose 
that a person has a strong view on a subject, believing that one alternative is 
very likely relative to alternatives. Suppose further that an inconclusive 
observation --- suggesting that one of the other alternatives is, in fact, the 
correct choice --- is presented to this person. Assuming that this person is 
rational, in which way would this evidence be assessed? 

If the person believes one of the alternatives is, \textit{a priori}, highly likely, 
then the uncertainty concerning this choice as measured by entropy is low. If 
this person has to change their point of view, then their posterior entropy will 
have to be raised before it can reduce again. However, raising entropy, generally, 
is counter to rational thinking. It follows that even if such inconclusive observations 
are presented repeatedly, if the prior view is highly skewed then it will take a 
long time for the posterior to change significantly. This is a feature of a rational 
assessment of an evidence that has been referred to as a ``tenacious Bayesian'' 
behaviour (Brody 2022). Intuitively, someone with very precise prior commitments 
is relatively impervious to contradictory evidence, and it will take a long time to 
change their mind.

With this in mind, let us consider what one might call the passive form of 
confirmation bias. This is concerned with how we accumulate evidence, and 
confirmation bias here refers to the effect that people interpret evidences on 
a subject of interest in such a way as to confirming their prior views on the 
matter. Influential empirical work by Lord, Ross and Lepper (1979) 
shows, in this context, that when people with diverging opinions on a 
subject matter (in this case the merits of capital punishment) are presented 
with identical evidences (articles to read) on the subject, afterwards 
the separation of their opinions increased.

\begin{figure}[t]
  \centering
       {\includegraphics[width=0.48\textwidth]{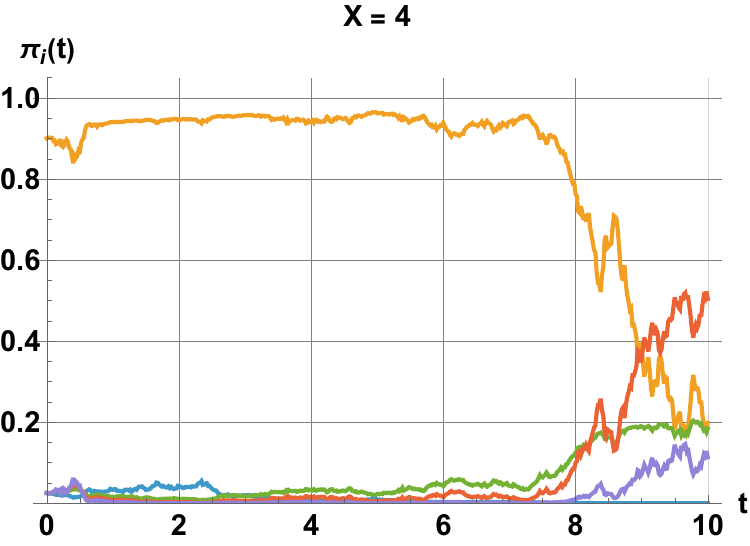}}\hfill
       {\includegraphics[width=0.48\textwidth]{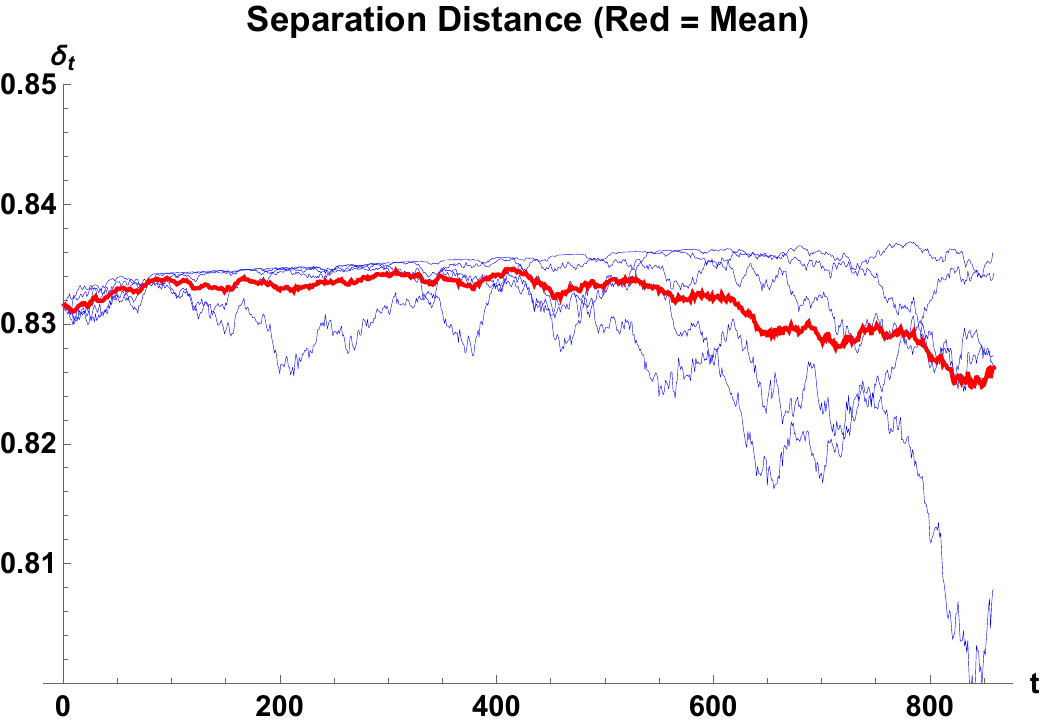}}\hfill
\caption{\textit{A tenacious Bayesian and separation distance}. Suppose 
that there are five alternatives labelled by $X=x_1,\ldots,x_5$. A Bayesian 
thinker has the prior view that ${\mathbb P}(X=x_2)=90\%$, and that all 
other alternatives are equally likely with the a prior of $2.5\%$. Suppose 
further that a noisy information time series $\xi_t=\sigma X t + \epsilon_t$ 
reveals that $X=x_4$ is in fact the correct alternative. Under a Bayesian 
updating, the view of this decision maker will hardly change for a long 
time, as illustrated in the left panel (the parameters are set at $x_k=k$ and 
$\sigma=0.4$ --- the posterior probability for the event $X=x_4$ is shown 
in orange, while that for the ``correct'' event $X=x_2$ is shown in red). 
Now suppose that there are two Bayesian decision makers, both with 
strong priors, but one with ${\mathbb P}(X=x_2)=96\%$ and the other 
with ${\mathbb P}(X=x_4)=96\%$. In each case, all other alternatives 
are viewed equally likely with the prior of $1\%$. However, a noisy 
information time series $\xi_t=\sigma X t + \epsilon_t$ suggests that the 
correct alternative is, say, $X=x_3$. In this case, their separation under 
Bayesian updating can increasing, as 
illustrated on the right panel (the parameters are set at $x_k=k$ and 
$\sigma=0.4$). Five sample paths are shown here, with the average in 
red. The average separation will indeed decrease, with a large 
sample size or with a very long time horizon. 
} 
\label{fig:z0} 
\end{figure}

Intuitively, if people are presented with the same evidence, then one would 
expect that --- under Bayesian updating --- their views will on average converge, 
rather than diverge. One would therefore conclude that the 
empirical behaviour observed in Lord \textit{et al}. (1979) indicates a sense 
of irrationality. 
The problem with this conclusion is that intuition alone is not a good 
scientific evidence. The intuition here is correct inasmuch as in a large 
sample-size limit --- or over a very long time period, where similar evidences 
are presented repeatedly --- diverging opinions will converge on average under 
Bayesian updating. 
However, over a shorter time horizon with finite sample size, the tenacious 
Bayesian behaviour tends to dominate. 

This fact can be illustrated by using a simple time-series model for evidence 
accumulation. Specifically, suppose that the object of interest is modelled by the 
random variable $X$, and that a time-series representing a sequence of 
observations is modelled by $\xi_t=\sigma X t + \epsilon_t$, where $\sigma$ 
is the rate at which reliable information is revealed through evidences and 
$\epsilon_t$ is a Brownian motion modelling noise. In other words, we 
decompose evidences in the ``signal-plus-noise'' form. In this case, as 
illustrated in Figure~\ref{fig:z0}, under Bayesian updating, the divergence 
of views as measured by the angular separation in Hilbert space of people 
with opposing priors has a tendency to increase in time (Brody 2022). In 
other words, confirmation bias of the type reported in Lord \textit{et al}. (1979) 
is consistent with standard adaptive Bayesian updating, and hence 
cannot be viewed as irrational. 

An important notion that underpins this observation is that when people with 
diverging views are presented with the same evidence, information contained 
in this evidence is handled differently by different people (Brody 2026). In 
information science, the amount of information contained in observed evidence 
about the unobserved and inferred variable of interest is measured in terms of 
mutual information (Gel’fand \& Yaglom 1957), which is observer dependent. A 
naive assertion that people respond differently to the same information 
therefore is not generally correct, because the same evidence does not imply 
the same mutual information, and just because people are given, say, the same 
article to read, it does not mean they aquire the same information. 

The notion that the information content of an observation is observer-dependent, 
along with the tenacious Bayesian argument, allows us to posit a passive form of 
confirmation bias in how people accumulate evidence. Needless to add, not every 
such confirmation bias observed empirically will admit rational explanation, 
but the analysis presented here shows that a rational behaviour based on Bayesian 
updating, without any \textit{ad hoc} modifications or constraints, is more 
versatile than what might have been expected from a naive interpretation of 
what a Bayesian updating ought to do.

\section*{D. Classical hypothesis testing for a binary decision} 
\label{app:1} 

Here we present the general theory of hypothesis testing for a decision among 
two alternatives. For simplicity we consider the following problem: We have a 
pair of coins, whose bias is known to be either $b_1$ or $b_2$. Hypothesis $k$ 
thus asserts that the bias of the coin is $b_k$. Let $p$ be the \textit{a priori} 
probability that the bias is $b_2$, and hence $1-p$ is the \textit{a priori} probability 
that the bias is $b_1$. Without loss we shall assume that $b_2>b_1$. 

The objective is to find the optimal strategy for choosing the hypothesis, subject 
to a given criterion (Berger 1985, DeGroot 1970). 
Suppose that we assign penalty $1$ for an incorrect choice, 
and $0$ for a correct choice. Such a penalty structure then determines the error 
probability. The task then is to determine a selection strategy that minimises the 
expected penalty. It should be intuitively clear that if a hypothesis has to be chosen 
without tossing the coin, then the optimal strategy that minimises the expected 
penalty is to choose hypothesis 2 if $p>1/2$, and hypothesis 1 otherwise. (In the 
degenerate case where $p=1/2$, either choice minimises the expected penalty.) 
The expected \textit{a priori} penalty $P_0^*(p)$ associated with the optimal 
strategy, when the coin has not been tossed, is therefore $P_0^*(p)=p$ if $p<1/2$ 
and $P_0^*(p)=1-p$ if $p\geq1/2$.  

Now suppose that the decision maker is allowed to toss the coin once, before 
choosing the hypothesis. The outcome will then provide partial information about 
the bias of the coin, from which the \textit{a priori} probability transforms into the 
\textit{a posteriori} probability. Hence we must consider a strategy that minimises 
the \textit{a posteriori} expectation of the penalty. Here the updating of the 
probability representing our perception about the hypotheses can be determined 
by use of the Bayes formula. Specifically, let $X$ denote the outcome of the first 
trial so that $X=1$ if the outcome is head, and $X=0$ if the outcome is tail. Then 
from the Bayes formula we deduce that the \textit{a posteriori} probability for 
hypothesis 2 is 
\begin{eqnarray} 
{\mathbb P}(H_2|X) = \frac{p\left( b_2X+(1-b_2)(1-X)\right)}{p\left( b_2X+(1-b_2)(1-X)\right) + (1-p)\left( b_1X+(1-b_1)(1-X)\right)} ,
\label{eq:6} 
\end{eqnarray} 
and the \textit{a posteriori} probability for hypothesis 1 is thus 
${\mathbb P}(H_1|X) =1-{\mathbb P}(H_2|X)$. 
Note that in the setup of the paper, the outcomes are modelled by the 
random variable $\xi=\pm1={\mathds 1}\{X=1\}-{\mathds 1}\{X=0\}$, 
in which case (\ref{eq:6}) is slightly modified to 
\begin{eqnarray} 
{\mathbb P}(H_2|\xi) = \frac{p\left( b_2(1+\xi)+(1-b_2)(1-\xi)\right)}
{p\left( b_2(1+\xi)+(1-b_2)(1-\xi)\right) + (1-p)\left( b_1(1+\xi)+(1-b_1)(1-\xi)\right)} .
\label{eq:6_2} 
\end{eqnarray} 
Rearranging the denominator and the numerator, we obtain 
\begin{eqnarray} 
\pi(\xi) = \frac{p\left[ 1+\xi(2b_2(\theta)-1) \right]}
{1+\xi\left[ 2p\,b_2(\theta)+2(1-p)b_1(\theta)-1 \right]} \, ,
\label{eq:z7} 
\end{eqnarray} 
which is the formula used in the paper. 
The optimal 
strategy that minimises the expected penalty, after a single coin tossing, 
is therefore to choose hypothesis 2 if 
${\mathbb P}(H_2|X) >1/2$, or equivalently, if 
\begin{eqnarray} 
p > \frac{b_1 X+(1-b_1)(1-X)}{(b_1+b_2)X+(2-b_1-b_2)(1-X)} . 
\label{eq:7} 
\end{eqnarray}
Otherwise, hypothesis 1 should be chosen. 

On a closer inspection, we see from (\ref{eq:7}) that when the outcome of the coin tossing 
is tail so that $X=0$, hypothesis 2 must be chosen if 
\begin{eqnarray} 
p > \beta = \frac{1-b_1}{(1-b_1)+(1-b_2)}  .
\label{eq:8} 
\end{eqnarray}
Conversely, for a head outcome we have $X=1$, so hypothesis 2 should be chosen if 
\begin{eqnarray} 
p > \alpha = \frac{b_1}{b_1+b_2}  .
\label{eq:9} 
\end{eqnarray}
However, because $b_2>b_1$ by assumption, 
it follows that $\beta>\alpha$. Therefore, if $p>\beta$ so that the 
\textit{a priori} probability is sufficiently high for hypothesis 2, then irrespective of the 
outcome of coin toss, the optimal strategy is to select hypothesis 2. Likewise, if 
$p<\alpha$ then hypothesis 1 must be chosen irrespective of the outcome. In the 
interval where $\alpha<p<\beta$, hypothesis 1 should be chosen when the outcome 
is $X=0$, whereas hypothesis 2 should be chosen when the outcome is $X=1$. 

The expected \textit{a posteriori} penalty, or the error probability, associated with 
the optimal strategy, which we shall denote by $P_1^*(p)$, when the coin has 
been tossed once, is therefore $P_1^*(p)=p$ if $p<\alpha$ and $P_1^*(p)=1-p$ if 
$p>\beta$, while for $\alpha<p<\beta$ we have 
\begin{eqnarray} 
P_1^*(p)&=& p\left( b_2\cdot0+(1-b_2)\cdot1\right) + (1-p)\left(b_1\cdot1+(1-b_1)\cdot0\right) 
\nonumber \\ &=& b_1+(1-b_1-b_2)p. 
\label{eq:10} 
\end{eqnarray}
This gives equation (4.3) of the paper. Hence in the less ambiguous regions 
$p<\alpha$ and $p>\beta$ we have $P_1^*(p) = P_0^*(p)$, but in the more 
ambiguous case $\alpha<p<\beta$ the expected cost is reduced by the 
information gained through the observation: $P_1^*(p)<P_0^*(p)$. 

If the decision maker is permitted to toss the coin more often, then the expected cost associated 
with an incorrect inference can be reduced further. Here we are interested in how rapidly the cost 
reduces as the number of trial is increased. For this purpose let us write $n$ for the number of 
times the coin is tossed before a decision is made, and $K$ for the number of heads in the 
$n$ trials. Thus $K$ is a random variable obeying a conditionally-binomial probability law: 
\begin{eqnarray}
{\mathbb P}(K=k) = p \binom{n}{k} b_2^k (1-b_2)^{n-k} + (1-p) 
\binom{n}{k} b_1^k (1-b_1)^{n-k} . 
\end{eqnarray} 
The reason for introducing the random variable $K$ is because in terms of $K$ the 
\textit{a posteriori} probabilities admit simpler expressions. Specifically, let us write 
$p_n={\mathbb P}(H_2|K)$ for the \textit{a posteriori} conditional probability that the 
bias of the coin is $b_2$, when the number of observed heads is $K$ after $n$ trials. 
Hence we have 
$p_0=p$, $p_1={\mathbb P}(H_2|X)$, and so on. Then from the Bayes formula we 
deduce that 
\begin{eqnarray}
p_n = \frac{p b_2^K (1-b_2)^{n-K}}{p b_2^K (1-b_2)^{n-K} + (1-p) b_1^K (1-b_1)^{n-K}} . 
\label{eq:14}
\end{eqnarray}
The optimal strategy after observing the outcomes of the $n$ coin tossing therefore is to 
choose hypothesis 2 if $p_n>1/2$, and hypothesis 1 otherwise. Equivalently, a short 
calculation shows that hypothesis 2 should be chosen if 
\begin{eqnarray}
K > \frac{\log\left[ \frac{p}{1-p} \left( \frac{1-b_2}{1-b_1}\right)^n\right]}
{\log\left[ \frac{b_1(1-b_2)}{b_2(1-b_1)}\right] } \equiv k_n . 
\end{eqnarray} 
The expected penalty (error probability) is therefore 
\begin{eqnarray}
P^*_n(p) &=& p\Big( 0\cdot {\mathbb P}(K>k_n|H_2) + 1 \cdot {\mathbb P}(K<k_n|H_2) 
\Big) \nonumber \\ && + (1-p) \Big( 0\cdot {\mathbb P}(K<k_n|H_1) + 1 \cdot 
{\mathbb P}(K>k_n|H_1) \Big) ,
\end{eqnarray}
which can be expressed explicitly in terms of cumulative binomial distribution 
functions. Specifically, defining 
\begin{eqnarray}
B_n(b,k) = \sum_{l=0}^{\lfloor k\rfloor} \binom{n}{l} b^l (1-b)^{n-l}  
\end{eqnarray}
for the cumulative Binomial distribution function, we find that 
\begin{eqnarray}
P^*_n(p) = p B_n(b_2,k_n) + (1-p) \Big( 1-B_n(b_1,k_n)  \Big) . 
\label{eq:18} 
\end{eqnarray}
The error probability thus obtained is therefore decreasing in $n$, but rather slowly, 
as indicated in Figure~1 (right panel) of the paper.

\section*{E. Quantum hypothesis testing for a binary decision} 
\label{app:2} 

Here we review the general theory of quantum hypothesis testing 
(Helstrom 1967, Holevo 1973, Stratonovic 1973, Yuen \textit{et al}. 
1975, Belavkin 1975, Helstrom 1976, Malley \& Hornstein 1993). The key 
difference here in the quantum inference from the classical counterpart in 
Section~D is that we are able to draw a conclusion based on a wide 
range of evidence --- not just from tossing the coin in the up-down directions but 
in every superposition of these two. In particular, as shown below, the choice of 
inference is dependent on the prior belief, in such a way that the higher the prior 
is for a given alternative, the closer the optimal inference will be to that alternative. 
Such a behaviour then leads on average to an exponential reduction in the cost 
associated with making a wrong choice, as the sample size is increased. 

In the case of quantum hypothesis testing for a binary decision, instead of given 
a coin whose bias can take one of the two possible values, we have an input 
quantum system whose state is given either by a density matrix ${\hat\rho}_2$, 
with the prior $p$, or by ${\hat\rho}_1$ with the prior $1-p$. We assume for 
simplicity that these two states belong to a two-dimensional Hilbert space so that 
we can compare the results with their classical counterparts discussed in the 
previous section. For convenience let us write $c_{ij}$ for the penalty associated 
with choosing hypothesis $H_i$ when $H_j$ is true. In our example we thus have 
$c_{ij}=1-\delta_{ij}$. 

The observational (quantum coin-tossing) strategy is represented in this case by a 
pair of projection operators ${\hat\Pi}_1$ and ${\hat\Pi}_2$ such that ${\hat\Pi}_1+
{\hat\Pi}_2={\mathds 1}$. Once the choice of the observables ${\hat\Pi}_1$ and 
hence ${\hat\Pi}_2$ are made, quantum hypothesis testing asserts that hypothesis 
$H_j$ should be chosen if the measurement outcome of the projector ${\hat\Pi}_j$ 
is positive. With these ingredients, and writing $p_2 = p$ and $p_1=1-p$ for the 
priors, the expected error probability (penalty) associated with an observational 
strategy $\{{\hat\Pi}_j\}$ is given by 
\begin{eqnarray}
P = \sum_{i,j=1}^2 p_j \, c_{ij} \, {\rm tr}\left( {\hat\Pi}_i {\hat\rho}_j \right) . 
\end{eqnarray} 
This follows because the probability of obtaining a positive outcome in the 
measurement ${\hat\Pi}_i$ when the state of the system is ${\hat\rho}_j$ is 
given by ${\rm tr}( {\hat\Pi}_i {\hat\rho}_j )$. 

The objective in quantum hypothesis testing therefore is to find a set of projectors 
$\{{\hat\Pi}_i^*\}$ such that the average error probability $P$ is minimised. Writing 
\begin{eqnarray}
{\hat R}_i = \sum_{j=1}^2 p_j \, c_{ij} \, {\hat\rho}_j 
\end{eqnarray}
for the Hermitian ``risk operator'', the error probabilitycan be written in the form 
\begin{eqnarray}
P = {\rm tr}\left( \sum_{i=1}^2 {\hat R}_i  \,  {\hat\Pi}_i \right) . 
\end{eqnarray} 
The optimality conditions on the measurements $\{{\hat\Pi}^*_i\}$ are known 
to be the self-adjointness of the operator 
${\hat R}_1{\hat\Pi}^*_1+{\hat R}_2{\hat\Pi}^*_2$ and the nonnegativity of the 
operators ${\hat R}_1 - ({\hat R}_1{\hat\Pi}^*_1+{\hat R}_2{\hat\Pi}^*_2)$ 
and ${\hat R}_2 - ({\hat R}_1{\hat\Pi}^*_1+{\hat R}_2{\hat\Pi}^*_2)$. Now a short 
calculation making use of the relation ${\hat\Pi}_1+{\hat\Pi}_2={\mathds 1}$ 
shows that 
\begin{eqnarray}
{\hat R}_1 - ({\hat R}_1{\hat\Pi}^*_1+{\hat R}_2{\hat\Pi}^*_2) = ({\hat R}_1 - {\hat R}_2)
{\hat\Pi}^*_2 = p_2 ({\hat\rho}_2-\gamma{\hat\rho}_1){\hat\Pi}^*_2 \geq {\mathbf 0}, 
\label{eq:22} 
\end{eqnarray}
where $\gamma=p_1/p_2$. Similarly we have 
\begin{eqnarray}
{\hat R}_2 - ({\hat R}_1{\hat\Pi}^*_1+{\hat R}_2{\hat\Pi}^*_2) = ({\hat R}_2 - {\hat R}_1)
{\hat\Pi}^*_1 = -p_2 ({\hat\rho}_2-\gamma{\hat\rho}_1){\hat\Pi}^*_1 \geq {\mathbf 0} . 
\label{eq:23} 
\end{eqnarray}
Hence if we write $\eta_i$ and $|\eta_i\rangle$ for the eigenvalues and eigenstates of 
the Hermitian matrix ${\hat\rho}_2-\gamma{\hat\rho}_1$, then from $p_2\geq0$ we 
have it from (\ref{eq:22}) and (\ref{eq:23}) that 
\begin{eqnarray}
\langle \eta_i|({\hat\rho}_2-\gamma{\hat\rho}_1){\hat\Pi}^*_2|\eta_i\rangle = 
\eta_i \langle \eta_i|{\hat\Pi}^*_2|\eta_i\rangle \geq 0 
\end{eqnarray}
and that 
\begin{eqnarray}
\langle \eta_i|({\hat\rho}_2-\gamma{\hat\rho}_1){\hat\Pi}^*_1|\eta_i\rangle = 
\eta_i \langle \eta_i|{\hat\Pi}^*_1|\eta_i\rangle \leq 0 . 
\end{eqnarray}
Now if $\eta_i<0$ then it must be $\langle \eta_i|{\hat\Pi}^*_2|\eta_i\rangle \leq 0$, but 
${\hat\Pi}^*_2$ is nonnegative, so it follows that for $\eta_i<0$, we have $\langle \eta_i| 
{\hat\Pi}^*_2|\eta_i\rangle=0$. Similarly, for $\eta_i>0$ it follows that $\langle \eta_i| 
{\hat\Pi}^*_1|\eta_i\rangle=0$. 
Therefore, the operator $\Pi_2^*$ projects onto the subspace of the Hilbert space spanned 
by the eigenvectors $|\eta_i\rangle$ corresponding to positive eigenvalues, and hence 
$\Pi_1^*$ onto the negative eigenspace. In other words, we have 
\begin{eqnarray}
{\hat\Pi}_1^* = \sum_i \Theta(-\eta_i) \, |\eta_i\rangle\langle\eta_i|  \quad {\rm and} \quad 
{\hat\Pi}_2^* = \sum_i \Theta(\eta_i) \, |\eta_i\rangle\langle\eta_i|, 
\end{eqnarray}
where $\Theta(u)$ denotes the Heaviside function: $\Theta(u)=0$ for $u<0$ and 
$\Theta(u)=1$ for $u>0$. 

The optimal error probability therefore is determined by 
$P^*={\rm tr}({\hat R}_1{\hat\Pi}_1^*+{\hat R}_2{\hat\Pi}_2^*)$. Substituting 
${\hat R}_1=p_2{\hat\rho}_2$ and ${\hat R}_2=p_1{\hat\rho}_1$, and using the 
relation ${\hat\Pi}_2^* ={\mathds 1}-{\hat\Pi}_1^*$, we thus find 
\begin{eqnarray}
P^* &=& {\rm tr}\left( p_2{\hat\rho}_2{\hat\Pi}_1^*+p_1{\hat\rho}_1{\hat\Pi}_2^*\right) 
\nonumber \\ &=&  {\rm tr}\left( p_1 {\hat\rho}_1 + p_2({\hat\rho}_2-\gamma{\hat\rho}_1) 
{\hat\Pi}_1^* \right) \nonumber \\ &=& p_1 + p_2 \sum_{\eta_i<0} \eta_i \, . 
\end{eqnarray}

Let us now consider a quantum analogue of the classical coin-tossing example. We 
consider the case whereby the state of a given quantum system (or possibly many 
copies of an identical system) is known to be either a pure state $|\psi_1\rangle$ 
with probability $p_1=1-p$ or another pure state $|\psi_2\rangle$ with probability 
$p_2=p$, where 
\begin{eqnarray}
|\psi_i\rangle = \cos\half \theta_i \, |0\rangle+\sin\half\theta_i \, |1\rangle \, .
\end{eqnarray} 
Equivalently, in a density matrix form we can write 
\begin{eqnarray}
{\hat\rho}_{i} = \left( \begin{array}{cc} 
\cos^2\frac{1}{2}\theta_i & \sin\frac{1}{2}\theta_i\,\cos\frac{1}{2}\theta_i \\ 
\sin\frac{1}{2}\theta_i\,\cos\frac{1}{2}\theta_i & \sin^2\frac{1}{2}\theta_i \end{array} \right) .
\label{eq:xxx} 
\end{eqnarray} 
This quantum ``coin'' therefore has the bias of $b_1=\sin^2\frac{1}{2}\theta_1$ or 
$b_2=\sin^2\frac{1}{2}\theta_2$. 

To identify the optimal detection strategy $\{{\hat\Pi}^*_i\}$ we need to determine the 
eigenstate of ${\hat\rho}_2-\gamma{\hat\rho}_1$ with the negative eigenvalue. To 
simplify the analysis we note that the eigenstates of ${\hat\rho}_2-\gamma{\hat\rho}_1$ 
are identical to those of its trace-free part 
\begin{eqnarray}
\left( \begin{array}{cc} 
z & x \\ x & -z \end{array} \right) := \half \left( \begin{array}{cc} 
\cos\theta_2-\gamma\cos\theta_1 & 
\sin\theta_2-\gamma\sin\theta_1 \\ 
\sin\theta_2-\gamma\sin\theta_1 & 
\cos\theta_2-\gamma\cos\theta_1 \end{array} \right) ,
\label{eq:xxxx} 
\end{eqnarray} 
whose eigenvalues are $\pm\sqrt{x^2+z^2}$. Expressing the eigenstate 
corresponding to the negative eigenvalue in the form 
\begin{eqnarray}
|\Pi_-^*\rangle = \cos\half \theta^* \, |0\rangle + \sin\half\theta^* \, |1\rangle , 
\end{eqnarray} 
we deduce that 
\begin{eqnarray}
\tan\half\theta^* = \frac{x}{z-\sqrt{x^2+z^2}} . 
\label{eq:32} 
\end{eqnarray} 
However, using a simple trigonometric identity 
\begin{eqnarray}
\tan\theta = \frac{2\tan\frac{1}{2}\theta}{1-\tan\frac{1}{2}\theta} 
\end{eqnarray}
we find from (\ref{eq:32}) that $\tan\theta^*=x/z$, and hence that 
\begin{eqnarray}
\theta^* = \tan^{-1} \left( 
\frac{p \sin\theta_2-(1-p)\sin\theta_1}{p \cos\theta_2-(1-p)\cos\theta_1} \right) 
\label{eq:34} 
\end{eqnarray} 
for the optimal detection angle. 

We see that (\ref{eq:34}) gives an alternative derivation of the argument 
presented in the paper to arrive at the optimal choice $\theta^*$, but 
how can this be used to derive an alternative representation 
\begin{eqnarray}
\pi^*(p) = \frac{1}{2}\left( 1+\xi\sqrt{1-4p(1-p)\cos^2\half\delta}\right) ,  
\label{eq:z10} 
\end{eqnarray} 
for the posterior? To see this, we substitute (\ref{eq:34}) in 
\begin{eqnarray}
b_k(\theta^*) = \cos^2 \half (\theta^*-\theta_k)  
\label{eq:z6}
\end{eqnarray}
and make use of the trigonometric identity 
\begin{eqnarray}
\cos\left( \tan^{-1}(z)-\theta_k\right) = \frac{1}{1+z^2}\,\left( \cos\theta_k - z 
\sin\theta_k\right) \, . 
\end{eqnarray}
Then for 
\begin{eqnarray}
z = \frac{p \sin\theta_2-(1-p)\sin\theta_1}{p \cos\theta_2-(1-p)\cos\theta_1}  
\end{eqnarray} 
we find 
\begin{eqnarray}
\frac{1}{1+z^2} = \frac{p \sin\theta_2-(1-p)\sin\theta_1}
{\sqrt{1-4p(1-p)\cos^2\frac{1}{2}\delta}} \, , 
\end{eqnarray} 
with $\delta=\theta_2-\theta_2$. Substituting this in (\ref{eq:z6}) and then 
the resulting expression in (\ref{eq:z7}), and eliminating the square root 
from the denominator, we arrive at (\ref{eq:z10}) after a short algebra.

\section*{F. Can a brain really explore superpositions?}
\label{app:3} 

In a typical introduction to quantum computing, the starting point of the 
discussion concerns the superposition principle of quantum mechanics. While a 
classical ``bit'' can only take two distinct values $0$ and $1$, a quantum bit, or a 
``qubit'', can take the form of an arbitrary superposition: 
$\cos\frac{1}{2}\theta|0\rangle
+\sin\frac{1}{2}\theta \, \re^{{\rm i}\phi}|1\rangle$ of the binary states $|0\rangle$ and 
$|1\rangle$ for $0\leq\theta\leq\pi$ and $0\leq\phi<2\pi$. Thus a quantum computer 
is a kind of a hybrid of analog and digital computers: the variables $(\theta,\phi)$ can 
be adjusted continuously, but when an output is read, it is discrete and takes on a 
binary value.  
On account of the existence of extra degrees of freedom offered by 
superpositions, there are certain tasks, such as the factorisation of an integer or 
the search of an unstructured database, for which a quantum computer can 
outperform its classical counterpart. Although these concepts are well 
documented in the literature, what is perhaps less appreciated is a result on 
hypothesis testing for which a quantum algorithm also outperforms its classical 
counterpart. 

A hypothesis testing concerns the determination of the validity of a given hypothesis. 
Take the simple example in a binary setting that is analogous to a classical bit: a coin 
is given, whose 
bias is known to be either $b_1$ (hypothesis 1) or $b_2$ (hypothesis 2). The 
\textit{a priori} probability for hypothesis 2 is $p$, and hence that for hypothesis 1 is 
$1-p$. As we have discussed in Section~D, 
classically, we can test these hypotheses by means of a series of 
measurements involving coin tossing. In particular, if the number of coin tossing is 
sufficiently large, then the frequency of getting the ``head'' outcome will be close to 
either $b_1$ or $b_2$, so we can be confident which hypothesis is the correct one. 

Letting $|1\rangle$ denote the head state and similarly $|0\rangle$ for the tail state, 
a quantum coin can be in a state of superposition $\cos\frac{1}{2}\theta|0\rangle
+\sin\frac{1}{2}\theta \, \re^{{\rm i}\phi}|1\rangle$. A binary quantum hypothesis 
testing is to determine if this quantum coin is in the state 
$\cos\frac{1}{2}\theta_1|0\rangle
+\sin\frac{1}{2}\theta_1 \, \re^{{\rm i}\phi_1}|1\rangle$, with bias $b_1=\sin^2
\frac{1}{2}\theta_1$,  or in the state $\cos\frac{1}{2}\theta_2|0\rangle
+\sin\frac{1}{2}\theta_2 \, \re^{{\rm i}\phi_2}|1\rangle$, with bias $b_2=\sin^2
\frac{1}{2}\theta_2$. But more importantly, \textit{a quantum coin can be 
tossed in the direction of an arbitrary superposition}. Although the outcome of each 
measurement is a binary number, the fact that we can explore all possible 
superpositions gives us an advantage in that a quantum hypothesis testing 
outperforms its classical counterpart in reducing the error probability. It follows 
that for a given level of 
error probability in choosing the hypothesis, a quantum algorithm requires a smaller 
sample size to reach that level. Alternatively stated, the error probability decreases 
exponentially in sample size in quantum hypothesis testing, in contrast to its classical 
counterpart. 

Our proposal here is based on the thesis that whenever there is an uncertainty among 
a given set of hypotheses or choices, the state of mind in relation to that set is in 
a superposition. The situation can be explained using 
the example of coin tossing. When a coin of bias $b$ is tossed but the outcome is not 
revealed, then the state of the coin can be described as either head with probability $b$ 
or tail with probability $1-b$. Such a state (of the classical coin) in quantum theory is 
described as a mixed state density matrix 
\begin{eqnarray}
{\hat\rho}_{\rm coin} = \left( \begin{array}{cc} b & 0 \\ 0 & 1-b \end{array} \right) . 
\label{eq:1.1} 
\end{eqnarray} 
However, if a person has to guess the outcome, then until a choice is made, the 
state of mind, we argue, is represented as a pure state 
\begin{eqnarray}
{\hat\rho}_{\rm mind} = \left( \begin{array}{cc} b & \sqrt{b(1-b)} \\ \sqrt{b(1-b)} & 1-b 
\end{array} \right) ,
\label{eq:1.2} 
\end{eqnarray} 
or equivalently in a state of superposition $\sqrt{b}|1\rangle+\sqrt{1-b}|0\rangle$. 
In other words, there is a discrepancy between the physical state of 
the coin, and the state of the mind of a person trying to guess the state of the coin. 

As we shall discuss below, classically, these two characterisations (mixed and 
pure) are merely alternative representations of the same state, and they are 
indistinguishable. That is, a pure state (\ref{eq:1.2}) that is in a superposition of 
two or more definite states is merely a representation of probability on Hilbert 
space, having no intrinsic significance. However, quantum mechanically, these 
two states are physically distinguishable. It follows that if these two states can 
be distinguished in cognitive science, then it follows that the notion of 
superposition is exploited by human brain, and this in turn will make our 
``rational'' explanation of confirmation bias plausible. With this in mind, let 
us describe in the Section below how a pure state and its associated mixed state 
can be distinguished in quantum theory, and suggest a thought experiment in 
cognitive science for testing the superposition principle.

\section*{G. Distinguishing pure and mixed states} 
\label{app:4} 

Consider a quantum system modelled on a two-dimensional Hilbert space, a generic 
element of which can be expressed in the form 
\begin{eqnarray}
|\psi\rangle = \cos\half \theta \, |0\rangle
+\sin\half\theta \, \re^{{\rm i}\phi} \, |1\rangle 
\end{eqnarray} 
with respect to a pair of orthonormal vectors $|0\rangle$ and $|1\rangle$. The vector 
$|\psi\rangle$ is understood to represent the \textit{state} of the system, which can be 
found to be in the state $|0\rangle$ with probability $\cos^2\half \theta$, and $|1\rangle$ 
with probability $\sin^2\half \theta$. An alternative 
way of expressing the state is in terms of a density matrix, which in the 
$\{|0\rangle,|1\rangle\}$-bases is given by 
\begin{eqnarray}
{\hat\psi} = |\psi\rangle\langle\psi| = \left( \begin{array}{cc} \cos^2\frac{1}{2}\theta & 
\sin\frac{1}{2}\theta\cos\frac{1}{2}\theta\, \re^{{\rm i}\phi}  \\ 
\sin\frac{1}{2}\theta\cos\frac{1}{2}\theta\, \re^{-{\rm i}\phi} & \sin^2\frac{1}{2}\theta 
\end{array} \right) . 
\label{eq:2} 
\end{eqnarray} 
For example, if $\theta=2\cos^{-1}\sqrt{b}$ and $\phi=0$ then (\ref{eq:2}) gives the 
state of mind (\ref{eq:1.2}) in the coin-tossing example. 

A typical observable ${\hat F}$ 
is represented in terms of a Hermitian matrix, such as 
\begin{eqnarray}
{\hat F} = f_0 |0\rangle\langle0| + f_1 |1\rangle\langle1| 
= \left( \begin{array}{cc} f_0 & 0 \\ 0 & f_1 \end{array} \right) ,
\label{eq:3} 
\end{eqnarray} 
where the two eigenvalues $f_0,f_1$ represent possible outcomes of a measurement of the 
observable represented by ${\hat F}$. If the state of the system is $|\psi\rangle$, and if 
${\hat F}$ is measured, then the probability of detecting the outcome $f_0$ is 
$\cos^2\frac{1}{2}\theta$, whereas that of detecting the outcome $f_1$ is 
$\sin^2\frac{1}{2}\theta$. In the former case, the state of the system after measurement 
will be $|0\rangle$, and similarly for the latter it will be $|1\rangle$. 

A state is said to be ``pure'' if its matrix representation corresponds to a projection onto 
a given state vector, and hence has the matrix rank one. More generally, a quantum 
state can be ``mixed'' and is expressible in the form of a mixed-state density matrix 
\begin{eqnarray}
{\hat \rho} = p |0\rangle\langle0| + (1-p) |1\rangle\langle1| 
= \left( \begin{array}{cc} p & 0 \\ 0 & 1-p \end{array} \right) . 
\label{eq:4} 
\end{eqnarray} 
If $\theta=2\cos^{-1}(\sqrt{p})$ and $\phi$ is arbitrary so that $|\cos\frac{1}{2}\theta|^2=p$ 
and $|\sin\frac{1}{2}\theta\,\re^{{\rm i}\phi}|^2=1-p$, then a pure state ${\hat\psi}$ (or 
equivalently a state vector 
$|\psi\rangle$) represents an indeterminate state that, upon measurement of an observable 
${\hat F}$, can reduce to a definite state $|0\rangle$ with probability $p$, and similarly to 
$|1\rangle$ with probability $1-p$. In contrast, a mixed state ${\hat\rho}$ is one for which 
the system is either in the state $|0\rangle$ with probability $p$ or in the state $|1\rangle$ 
with probability $1-p$. For example, if the system is initially in the state $|\psi\rangle$ and 
a measurement of the observable ${\hat F}$ is performed but the outcome not recorded, 
then after the measurement the state of the system is ${\hat\rho}$. That is, after the 
measurement, the state of the system is known to be in $|0\rangle$ or in $|1\rangle$, but 
the observer does not know which. This is the essential difference between a pure 
state and a mixed state. In one case, the state is indefinite, representing in some sense intrinsic probabilities for the undecided, whereas in the other case the state 
is definite but it is unknown to an external observer, thus representing in some 
sense extrinsic probabilities for the unknown. 

To examine the difference between the pure state ${\hat\psi}$ and the associated mixed 
state ${\hat\rho}$ let us consider moments of the observable ${\hat F}$, because in 
classical probability the information about a state is encoded in its moments. A short 
calculation then shows that the moments are identical in both cases: ${\rm tr}({\hat\rho}
{\hat F}^n)={\rm tr}({\hat\psi}{\hat F}^n)=p f_0^n + (1-p)f_1^n$. It follows that if every 
observable is expressible in the form (\ref{eq:3}), and hence they are commutative, then 
a pure state $|\psi\rangle$ and the associated mixed state ${\hat\rho}$ are indistinguishable 
and hence equivalent. Because in classical physics (or in classical probability) all 
observables are commutative, one can work either with probabilities as represented by 
a density matrix ${\hat\rho}$, or with square-root probabilities as represented by a pure 
state $|\psi\rangle$ --- both representations are equivalent. That is to say, in classical 
probability, there is no distinction between an intrinsic probability (for example, the state 
of a tossed coin still in mid air) and an extrinsic probability (for example, the state 
of an already tossed coin whose outcome is hidden). 

Quantum mechanically, on the other hand, not all observables are commutative. 
Take, for example, another observable ${\hat G}$ defined by 
\begin{eqnarray}
{\hat G} = g |0\rangle\langle1| + {\bar g} |1\rangle\langle0| 
= \left( \begin{array}{cc} 0 & g \\ {\bar g} & 0 \end{array} \right) ,
\label{eq:5} 
\end{eqnarray} 
where $g$ is an arbitrary complex number and ${\bar g}$ its complex conjugate. 
A measurement of ${\hat G}$ will yield one of the two possible outcomes 
$\pm\sqrt{{\bar g}g}$ corresponding to the two eigenvalues of ${\hat G}$. In this 
case, already the first moment of ${\hat G}$ is different between the two 
states: We have ${\rm tr}({\hat\rho}{\hat G})=0$, whereas for the pure state, 
${\rm tr}({\hat\psi}{\hat G})=\sqrt{p(1-p)}(g\re^{{\rm i}\phi}+{\bar g}\re^{-{\rm i}\phi})$. 
Hence the two states ${\hat\rho}$ and ${\hat\psi}$ 
are physically distinct and statistically distinguishable, when 
there are incompatible observables. Indeed, the distinguishability of a pure state 
and its associated mixed state is one of the important features of quantum 
probability rule that has no analogue in classical probability. 

In the case of human behaviour, literature in the field of quantum cognition has 
provided empirical observations that strongly suggest that many human decisions 
or propositions are not compatible. This motivates us to consider the following 
thought experiment. Consider, say, two binary choices, labelled by $A$ and $B$. 
These choices must be selected carefully to avoid the ones for which many people 
would have encountered or considered previously, for, otherwise, the state of mind 
would already be in a mixed state in relation to these choices before the 
experiment. Putting it differently, these choices must be somewhat unexpected. 
Additionally, these choices must be those that are expected to be incompatible 
from other cognitive experiments. The notion of incompatibility must be clarified 
here because many people mistakenly assume that it means the outcome of 
$A$ affecting the outcome of $B$. Clearly, such a causal influence can be seen 
in classical probability, whenever $A$ and $B$ are not statistically independent. 
The incompatibility here refers to the fact that the act of inferring $A$, and not 
the outcome, influencing the statistics of $B$. 

Now for a given set of people, half of them will be asked to make the choice $A$ 
followed by choice $B$. The idea is that having made the choice $A$, their states 
of mind are in a mixed state in the reference frame of choice $A$, when choice 
$B$ is assessed. The other half will be asked to simply make the choice $B$. The 
discrepancy in the statistics for the outcome of $B$ will then suggest that states 
of mind of the latter group, in the reference frame of choice $A$, are not mixed. 

Experiments analogous to the one considered here have indeed been performed, 
for instance in Busemeyer \& Wsng (2015), but those experiments are primarily 
designed to 
demonstrate the existence of incompatibility between $A$ and $B$. One can, 
however, turn the argument around to show that these experiments indeed show 
not only the incompatibility of $A$ and $B$, but also the existence of a pure state 
of mind that is in a superposition of definite choices. This follows from the fact 
that if the initial uncertainty associated with choice $A$ were arising from a 
mixed state (that is, arising from a classical uncertainty), then even if $B$ is 
incompatible with $A$, the statistics of the two groups cannot differ significantly. 
That the state of mind can be in a coherent superposition is significant because 
it provides us with an extra resource, which in turn will influence our strategies 
for hypothesis testing.

\end{document}